\documentclass[english,useAMS, usenatbib]{mn2e}
\usepackage[T1]{fontenc}
\usepackage[latin1]{inputenc}
\usepackage{times}
\usepackage{graphicx}
\usepackage{amssymb}
\usepackage{babel}

\makeatletter

\IfFileExists{url.sty}{\usepackage{url}}
                      {\newcommand{\url}{\texttt}}
\providecommand{\tabularnewline}{\\}

\newcommand{\aap}{A\&A}
\newcommand{\aaps}{A\&AS}
\newcommand{\apj}{ApJ}
\newcommand{\apjl}{\apj}

\newcommand{\aj}{AJ}
\newcommand{\mnras}{MNRAS}
\newcommand{\apjs}{ApJS}
\newcommand{\aapr}{A\&AR}
\newcommand{\pasj}{PASJ}

\newcommand{\kmps}{\mathrm{km~s^{-1}}}
\newcommand\ion[2]{#1$\,${\sc {#2}}}   
\newcommand{\Kelvin}{\mathrm{K}}

\makeatother
\begin{document}
\title[On the formation of H$\alpha$ around classical T Tauri
  stars]{On the formation of H$\alpha$ line emission around classical
  T Tauri stars}

\author[R. Kurosawa et\,al.]{Ryuichi
  Kurosawa$^1$\thanks{E-mail:rk@astro.ex.ac.uk}, Tim J. Harries$^1$
  and Neil H. Symington$^2$\\ $^1$School of Physics, University of Exeter, Stocker Road, Exeter EX4~4QL. \\ $^2$School of Physics and Astronomy, University of St. Andrews, North Haugh, St. Andrews, Fife, KY16 9SS.}

\date{Dates to be inserted}

\pagerange{\pageref{firstpage}--\pageref{lastpage}} \pubyear{2005}

\maketitle

\label{firstpage}

\begin{abstract}

We present radiative transfer models of the circumstellar environment
of classical T~Tauri stars, concentrating on the formation of the
H$\alpha$ emission. The wide variety of line profiles seen in
observations are indicative of both inflow and outflow, and we
therefore employ a circumstellar structure that includes both
magnetospheric accretion and a disc wind.  We perform
systematic investigations of the model parameters for the wind and the
magnetosphere to search for possible geometrical and physical
conditions which lead to the types of profiles seen in
observations. We find that the hybrid models can reproduce the wide
range of profile types seen in observations, and that the most common
profile types observed occupy a large volume of parameter
space. Conversely, the most infrequently observed profile morphologies
require a very specific set of models parameters.  We find our model
profiles are consistent with the canonical value of the 
mass-loss rate to mass-accretion rate ratio ($\mu=0.1$) found in
earlier magneto-hydrodynamic calculations and observations, but the
models with $0.05<\mu<0.2$ are still in accord with observed
H$\alpha$ profiles.  We investigate the
wind contribution to the line profile as a function of model parameters, and
examine the reliability of H$\alpha$ as a mass accretion diagnostic.
Finally, we examine the
H$\alpha$ spectroscopic classification used by Reipurth et.~al, and
discuss the basic physical conditions that are required to reproduce
the profiles in each classified type.

\end{abstract}


\section{Introduction }

\label{sec:Introduction}

T~Tauri stars (TTS) are young ($\gtrsim 3\times10^{6}\,\mathrm{yrs}$,
\citealt{appenzeller:1989}) low-mass objects, and are the progenitors
of solar-type stars. Classical T~Tauri stars (CTTS) exhibit strong
H$\alpha$ emission, and typically have spectral types of F--K. Some of
the most active CTTS show emission in higher Balmer lines and metal
lines (e.g., \ion{Ca}{ii}~H and K). They also exhibit excess continuum
flux in the ultraviolet (UV) and infrared (IR).  Their spectral energy
distribution and polarisation data suggest the presence of
circumstellar discs, which play an important role in regulating
dynamics of gas flows around CTTS (e.g. \citealt{camenzind:1990}).

Many observational studies (e.g., \citealt{herbig:1962};
\citealt{edwards:1994}; \citealt{kenyon:1994};
\citealt*{reipurth:1996}; \citealt{alencar:2000}) of CTTS line
profiles have revealed evidence for both outward wind flows and inward
accretion flows, characterised by the blue-shifted absorption features
in H$\alpha$ profiles and redshifted inverse P~Cygni (IPC)
profiles respectively. Typical mass-loss rates of CTTS are about
$10^{-9}$ to $10^{-7}\,\mathrm{M_{\sun}\,
yr^{-1}}$ (e.g., \citealt{kuhi:1964}; \citealt{edwards:1987};
\citealt*{hartigan:1995}), and the mass-accretion rates are also about
$10^{-9}$ to $10^{-7}\,\mathrm{M_{\sun}\,
yr^{-1}}$ (e.g., \citealt{kenyon:1987}; \citealt*{bertout:1988};
\citealt{gullbring:1998}).  Recent H$\alpha$ spectro-astrometric
observations by \citet*{takami:2003} provide indirect evidence for the
presence of bipolar and monopolar outflows down to $\sim
1\,\mathrm{au}$ scale (e.g.\,CS~Cha and RU~Lup).  Similarly, ESO VLT
observations using high-resolution ($R=50\,000$) two-dimensional
spectra of edge-on CTTS (HH~30, HK~Tau~B, and HV~Tau~C) by
\citet{appenzeller:2005} show extended H$\alpha$ emission in the
direction perpendicular to the obscuring circumstellar disc,
suggesting the presence of the bipolar outflows. On an even larger
scale, \emph{HST} observations of HH30 \citep{burrows:1996} trace the
jet to within $\lesssim30$~au of the star. The jet has a cone shape
with an opening angle of $3^{\circ}$ between 70 and 700 au
\citep{koenigl:2000}. \citet{alencar:2000} found about 80 per cent of
their sample (30 CTTS) show blue-shifted absorption components in at
least one of the Balmer lines and Na~D.

In the currently favoured model of accretion in CTTS, the accretion
disc is disrupted by the magnetosphere, which channels the gas from
the disc onto the stellar surface (e.g., \citealt{uchida:1985};
\citealt{koenigl:1991}; \citealt{cameron:1993};
\citealt{shu:1994}). This picture is supported by recent measurements
of strong ($\sim 10^{3}\mathrm{G}$) magnetic fields in CTTS (e.g.,
\citealt{johns-krull:1999}; \citealt{symington:2005b}) and by
radiative transfer models which reproduce the gross characteristics of
observed profiles for some CTTS (\citealt*{muzerolle:2001}). In
particular the magnetospheric accretion (MA) model explains blue-ward
asymmetric emission line profiles as resulting from the partial
occultation of the flow by the stellar photosphere and by the inner
part of the accretion disc, while inverse 
P~Cygni profiles in the MA model result from inflowing material at
near free-fall velocities seen projected against hotspots on the
stellar surface.

Despite these successes, the overwhelming observational evidence for
outflows in CTTS suggests that the MA model is only one component
of a complex circumstellar environment. Clearly, one must include the
contribution of any wind/jet flow if one wishes to both accurately
predict the mass-accretion rate and also determine the mass-loss rate
of a CTTS via emission profile modelling. The first attempt in this
direction was made by \citet{alencar:2005} who demonstrated that the
observed H$\alpha$, H$\beta$ and Na~D lines of RW~Aur are better
reproduced by the radiative transfer model which included a collimated
disc-wind arising from near the inner edge of the accretion disc.

The main aim of this paper is to find a simple kinematic model which
can reproduce the wide variety of the observed profiles, and to
perform empirical studies of line formation in an attempt to place
morphological classification schemes on a firmer physical footing. We
will also discuss whether our model is consistent with some
predictions made by recent (magneto-hydrodynamics) MHD studies i.e.~
$\mu=\dot{M}_{\mathrm{wind}}/\dot{M}_{\mathrm{acc}}\approx0.1$
(e.g.~\citealt{koenigl:2000}).

In Section~\ref{sec:model-configuration}, the model assumptions and
the basic model configurations are presented. We discuss the radiative
transfer model used to compute the profiles in
Section~\ref{sec:radiative-transfer-model}, and the results of model
calculations are given in Section~\ref{sec:Results}.  We discuss our
results in the context of Reipurth's classification scheme in
Section~\ref{sec:Discussion}, and our summary and conclusions are
presented in Section~\ref{sec:Conclusions}.


\section{Model configuration}

\label{sec:model-configuration}

\begin{figure*}

  \begin{center}
    
    \includegraphics[%
      scale=0.6]{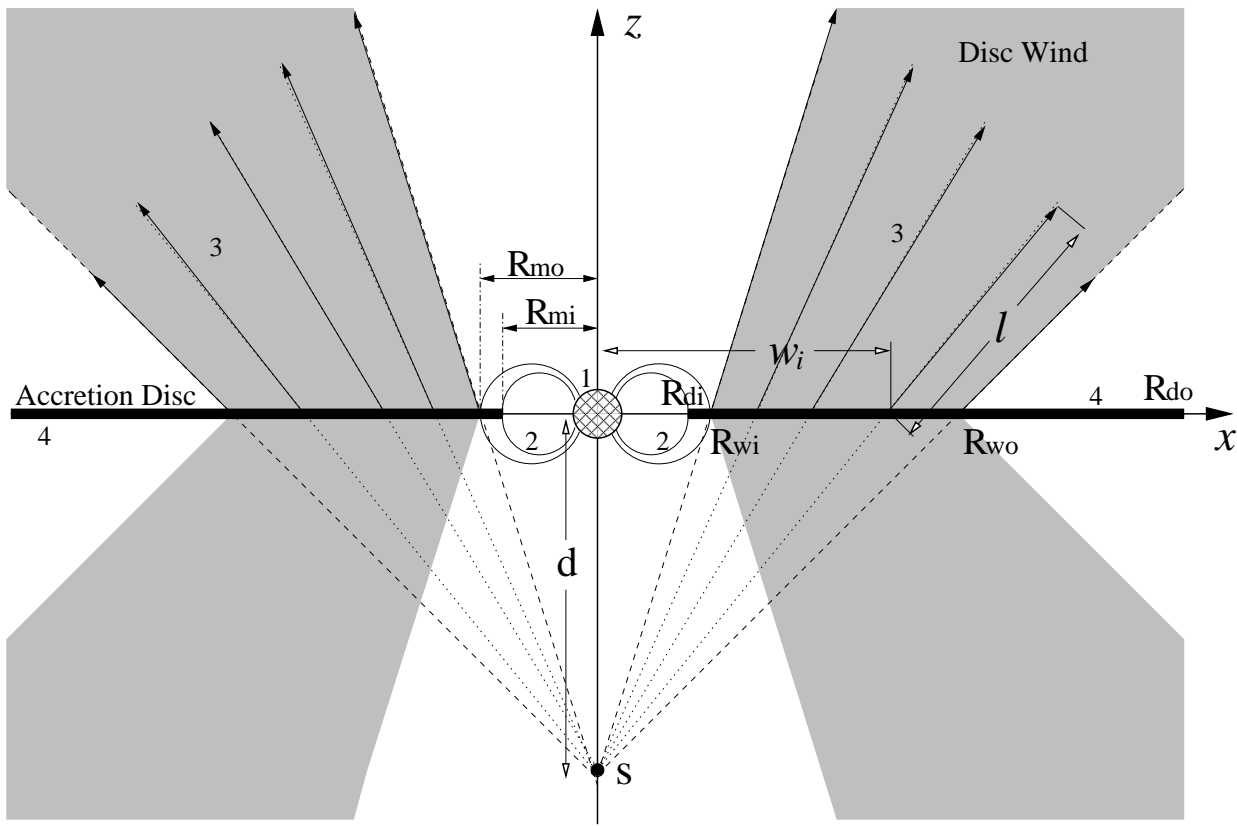}
    
  \end{center}

  \caption{Basic model configuration of the disc-wind-magnetosphere
    hybrid model. The system consist of four components: (1)~the
    continuum source located at the origin of the
    cartesian coordinates $\left(x,y,z\right)$ -- the $y$-axis is into
    the paper, (2)~the magnetospheric accretion flow, (3)~the disc
    wind, and (4)~the accretion disc. The disc wind originates from
    the disc surface between $w_{i}=R_{\mathrm{wi}}$ and
    $w_{i}=R_{\mathrm{wo}}$ where $w_{i}$ is the distance from the $z$
    axis on the equatorial plane.  The wind source
    points ($S$), from which the stream lines diverge, are placed at
    distance $d$ above and below the star. The degree of wind
    collimation is controlled by changing the value of $d$.}

  \label{fig:config_discwind}

\end{figure*}

In order to understand how the different parts of the CTTS
circumstellar environment contribute to the formation of H$\alpha$,
the model space is divided into four different regions: (1)~a central
continuum source, (2)~the magnetospheric accretion flow, (3)~the wind
outflow, and (4)~the accretion disc. Fig.~\ref{fig:config_discwind}
depicts the relative location of the regions in the model space.  The
density is assumed to be rotationally symmetric around the
$z$-axis. The innermost radius of the magnetosphere at the equatorial
plane coincides with the the inner radius of the accretion disc. From
the innermost part of the accretion disc, the gas falls freely, moving
along the magnetic field onto the surface of the star.  In the
following subsections we describe the details of the model components.

\subsection{The continuum source}

\label{sub:Continuum-Source}

We adopt stellar parameters of a typical
classical T~Tauri star for the central continuum source, i.e. radius
($R_{*}$), mass ($M_{*}$), and effective temperature of the photosphere
($T_{\mathrm{ph}}$) are $2\, \mathrm{R_{\sun}}$, $0.5\,
\mathrm{M_{\sun}}$, and $4000\,\mathrm{K}$ 
respectively. The model atmosphere of \citet{kurucz:1979} with
$T_{\mathrm{ph}}=4000\,\mathrm{K}$ and $\log g_{*}=3.5$ (cgs) defines
the photospheric contribution to the continuum flux. The
parameters are summarised in Table~\ref{tab:std_parameters}.

An additional continuum source is considered for models which include 
magnetospheric accretion: as the infalling gas approaches the stellar
surface, it decelerates in a strong shock, and is heated to $\sim
10^{6}\,\mathrm{K}$.  The X-ray radiation produced in the shock will
be absorbed by the gas locally, and re-emitted as optical and UV light 
(\citealt{calvet:1998}; \citealt{gullbring:2000}). This
will create hot rings where the magnetic field intersects with the
surface. We assume that the free-falling kinetic energy is thermalized
in the radiating layer, and is re-emitted as blackbody radiation with
a single temperature. With the parameters of the magnetosphere and the
star given above (Table~\ref{tab:std_parameters}), about $8$~per~cent
of the surface is covered by the hot rings. 
In our model, the fraction of hot-ring coverage area is
determined only by the geometry of the magnetosphere
(Section~\ref{sub:Magnetosphere}). \citet{gullbring:2000} estimated the
fractions of the hot-ring coverage area for three intermediate CTTS by
using the observed UV and optical spectra. They found the range of
fraction to be 0.3--3.0~per~cent.  This is smaller than the value adopted
here; however, even if we use 1~per~cent hot-ring area
coverage model (by changing the width of the magnetosphere), we
find the line shapes of H$\alpha$ do not change significantly (see
also Fig.~5 of \citealt{muzerolle:2001}).

If the mass-accretion rate
is $10^{-7}\,\mathrm{M_{\sun}\, yr^{-1}}$, the ratio of the accretion
luminosity to the photospheric luminosity is about $0.5$, and the
corresponding temperature of the hot rings is about
$6400\,\mathrm{K}$.  The luminosity ratio and the temperature depend
on an adopted mass-accretion rate. The continuum emission from the hot
rings is taken into account when computing the line profiles.

\begin{table*}

\caption{Summary of the reference classical T~Tauri star model
  parameters. Note that a typical mass-accretion rate of CTTS is $\sim 3
  \times 10^{-8}\,\mathrm{M_{\sun}\, yr^{-1}}$
  (e.g. \citealt{gullbring:1998}; \citealt{calvet:2004}); however, a
  slightly higher mass-accretion rate is adopted here to demonstrate the 
  effect of line-broadening more clearly (c.f. Section~\ref{sub:model-acc}). }
\label{tab:std_parameters}

\begin{center}

\begin{tabular}{lccccccccc}
\hline 
Parameters&
$R_{*}$&  
$M_{*}$&
$T_{\mathrm{ph}}$&
$R_{\mathrm{mi}}$&
$R_{\mathrm{mo}}$&
$\dot{M}_{\mathrm{acc}}$&
$\dot{M}_{\mathrm{wind}}$&
$R_{\mathrm{di}}$&
$R_{\mathrm{do}}$\tabularnewline
&
$\left[\mathrm{R_{\sun}}\right]$&
$\left[\mathrm{M_{\sun}}\right]$&
$\left[\mathrm{K}\right]$&
$\left[R_{*}\right]$&
$\left[R_{*}\right]$&
$\left[\mathrm{M_{\sun}}\,\mathrm{yr^{-1}}\right]$&
$\left[\mathrm{M_{\sun}}\,\mathrm{yr^{-1}}\right]$&
$\left[R_{*}\right]$&
$\left[\mathrm{au}\right]$\tabularnewline
\hline
&
$2.0$&
$0.5$&
$4000$&
$2.2$&
$3.0$&
$10^{-7}$&
$10^{-8}$&
$2.2$&
$100$\tabularnewline
\hline
\end{tabular}

\end{center}

\end{table*}

\subsection{The magnetosphere}

\label{sub:Magnetosphere}

We adopt the MA flow model of \citet*{hartmann:1994},
as done  by \citet{muzerolle:2001} and by \citet*{symington:2005},
in which the gas accretion on to the stellar surface from the innermost
part of the accretion disc occurs through a dipolar stellar magnetic
field. The magnetic field is assumed to be so strong that the gas
flow does not affect the underlying magnetic field itself. As shown
in Fig.~\ref{fig:config_discwind} (region~2), the innermost radius ($R_{\mathrm{mi}}$)
of the magnetosphere at the equatorial plane ($z=0$) is assigned
to be the same as the inner radius ($R_{\mathrm{di}}$) of the accretion
disc where the flow is truncated. In our models, $R_{\mathrm{mi}}$
and the outer radius ($R_{\mathrm{mo}}$) of the magnetosphere (at
the equatorial plane) are set to be $2.2\, R_{*}$ and $3.0\, R_{*}$
respectively. The geometry of the magnetic field/stream
lines is fixed for all calculations. We note that this magnetospheric geometry
is identical to the {}``small/wide'' model of \citet{muzerolle:2001}. 

The magnetic field and the gas stream lines are assumed to have the
following simple form 
\begin{equation}
      r=R_{\mathrm{m}}\sin^{2}\theta
  \label{eq:dipole}
\end{equation}
\citep*[see][]{ghosh:1977} where $r$, and $\theta$ are the radial and
the polar component of a field point position vector in the accretion
stream.  $R_{\mathrm{m}}$ is the radial distance to the field
line at the equatorial plane ($\theta=\pi/2$), and its value is
restricted between $R_{\mathrm{mi}}$ and 
$R_{\mathrm{mo}}$ (Fig.~\ref{fig:config_discwind}). 
Using the field geometry above and conservation of
energy, the velocity and the density of the accreting gas along the
stream line are found as in \citet{hartmann:1994}.

The temperature structure of the magnetosphere used by
\citet{hartmann:1994} is adopted here. They computed the temperature,
assuming a volumetric heating rate which is proportional to $r^{-3}$,
by solving the energy balance of the radiative cooling rate of
\citet{hartmann:1982} and the heating rate \citep{hartmann:1994}.
\citet{martin:1996} presented a self-consistent determination of the
thermal structure of the inflowing gas along the dipole magnetic field
(equation~\ref{eq:dipole}) by solving the heat equation coupled to the
rate equations for hydrogen. He found that main heat source is
adiabatic compression due to the converging nature of the flow, and
the major contributors to the cooling process are bremsstrahlung
radiation and line emission from Ca~II and Mg~II ions. 
\citet*{muzerolle:1998} found that the line profile models computed
according to the temperature structure of \citet{martin:1996} do not
agree with observations, unlike profiles based on the (less
self-consistent) \citet{hartmann:1994} temperature distribution.  It
is clear that the temperature structure of the magnetosphere is still
a large source of uncertainty in the accretion model; here we adopted
the model of \citet{hartmann:1994}.

\subsection{The disc wind}
\label{sub:model-disc-wind}

The magneto-centrifugal wind paradigm, first proposed by
\citet{blandford:1982}, has been often used to model the large-scale
wind structure of T Tauri stars, or the observed optical jets
(e.g.~HH~30 jet by \citealt{burrows:1996}; \citealt{ray:1996}). The
launching of the wind from a Keplerian disc is typically done by
treating the equatorial plane of the disc as a mass-injecting boundary
condition (e.g., \citealt{shu:1994}; \citealt{ustyugova:1995};
\citealt{ouyed:1997}; \citealt{krasnopolsky:2003}).  Depending on the
location of the open magnetic fields anchored to the disc, two
different types of winds are produced. If the field is constrained to
be near the co-rotation radius of the stellar magnetosphere, an
{}``X-wind'' \citep{shu:1994} is produced. If the open field lines are
located in a wider area of the disc, a {}``disc-wind'' similar to that
of \citet{koenigl:2000} is produced (see also
\citealt{krasnopolsky:2003}).  Recent reviews on the jet/wind-disc
connection can be found in \citet{koenigl:2000} and
\citet{pudritz:2005}. Clearly there are several alternative outflow
scenarios. Here we adopt a simple kinematical model, based on the
disc-wind paradigm, which broadly represents the results of MHD
simulations,

\citet*{knigge:1995} introduced the {}``split-monopole'' kinematic
disc-wind model in their studies of the UV resonance lines formed
in the winds of cataclysmic variable stars. Their formalism provides
a simple parameterisation of a disc wind that has similar properties 
to those found by  MHD modelling. In this model, the outflow
arises from the surface of the rotating accretion disc, and has a
biconical geometry. The specific angular momentum is assumed to be
conserved along a stream line, and the poloidal velocity component
is assumed to be simply a radial from {}``sources'' vertically displaced 
from the central star.  Here we briefly describe the disc-wind model, and
readers are referred to \citet{knigge:1995} and \citet{long:2002} for
details.

The four basic parameters of the model are: (1)~the mass-loss rate,
(2)~the degree of the wind collimation, (3)~the velocity gradient, and
(4)~the wind temperature. The basic configuration of the disc-wind
model is shown in Fig.~\ref{fig:config_discwind}. The disc wind
originates from the disc surface, but the {}``source'' point ($S$),
from which the stream lines diverge, are placed at distance $d$ above
and below the centre of the star.  The angle of the mass-loss
launching from the disc is controlled by changing the value of
$d$. The mass-loss launching occurs between $R_{\mathrm{wi}}$ and
$R_{\mathrm{wo}}$ where the former is set to be equal to the outer
radius of the magnetosphere ($R_{\mathrm{mo}}$) and the latter is set
to $1$~au,  as in \citet{krasnopolsky:2003}.

The local mass-loss rate per unit area ($\dot{m}$) is assumed to
be proportional to the mid-plane temperature of the disc, and is a
function of the cylindrical radius $w=\left(x^{2}+y^{2}\right)^{1/2}$, i.e.
\begin{equation}
  \dot{m}\left(w\right) \propto
       \left[ T\left(w\right) \right]  ^{4\alpha}\,\,.\label{eq:discwind-massloss-temp}
\end{equation}
The mid-plane temperature of the disc is assumed to be expressed
as a power-law in $w$; thus, $T\propto w^{q}$. Using this in the
relation above, one finds 
\begin{equation}
  \dot{m}\left(w\right)\propto w^{p}
  \label{eq:discwind-massloss-w}
\end{equation}
where $p=4\alpha\times q$. The index of the mid-plane temperature power
law is adopted from the dust radiative transfer model of
\citet{whitney:2003a} who found the innermost part of the accretion
disc has $q=-1.15$.  In order to be consistent with the collimated
disc-wind model of \citet{krasnopolsky:2003} who used $p=-7/2$, the
value of $\alpha$ is set to 0.76. The constant of proportionality in
equation~\ref{eq:discwind-massloss-w} is found by integrating
$\dot{m}$ from $R_{\mathrm{wi}}$ to $R_{\mathrm{wo}}$, and 
normalising the value to the total mass-loss rate
$\dot{M}_{\mathrm{wind}}$.

The azimuthal/rotational component of the wind velocity
$v_{\phi}\left(w,z\right)$ is computed from the Keplerian rotational
velocity at the emerging point of the stream line
i.e. $v_{\phi}\left(w_{i},0\right)=\left(GM_{*}/w_{i}\right)^{1/2}$
where $w_{i}$ is the distance from the rotational axis ($z$) to the
emerging point on the disc, and by assuming the conservation of the
specific angular momentum along a stream line:
\begin{equation}
  v_{\phi}\left(w,z\right)\, = 
  v_{\phi}\left(w_{i},0\right)\,\left(\frac{w_{i}}{w}\right)\,\,.
  \label{eq:discwind-toroidal-velocity}
\end{equation}
Based on the canonical $\beta$ velocity law of hot stellar winds
(c.f.~\citealt*{castor:1975}), the poloidal component of the wind
velocity ($v_{p}$) parameterised as:
\begin{equation}
  v_{p}\left(w_{i},l\right) = c_{\mathrm{s}}\left(w_{i}\right) +
  \left[f\, v_{\mathrm{esc}}-c_{\mathrm{s}}\left(w_{i}\right)\right]\left(1-\frac{R_{s}}{l+R_{s}}\right)^{\beta}
  \label{eq:discwind-poloidal-velocity}
\end{equation}  
where $c_{\mathrm{s}}$, $f$, and $l$ are the sound speed at the wind
launching point on the disc, the constant scale factor of the
asymptotic terminal velocity to the local escape velocity (from the
wind emerging point on the disc), and the distance from the disc
surface along stream lines respectively. $R_{s}$ is the wind scale
length, and its value is set to 10~$R_{\mathrm{wi}}$ by following
\citet{long:2002}. 

\begin{figure*}
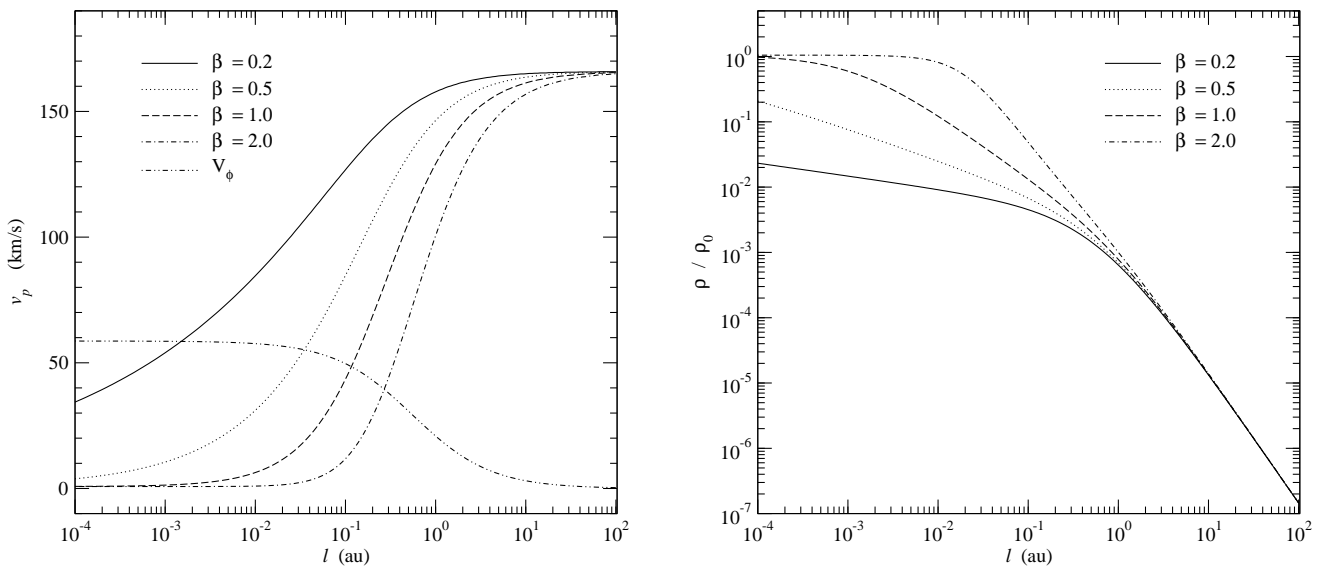


  \begin{center}
    
    \begin{tabular}{cc}
      \includegraphics[%
	clip,
	scale=0.59]{fig02a.eps}~~~~~~&
      \includegraphics[%
	clip,
	scale=0.59]{fig02b.eps}\tabularnewline
    \end{tabular}
    
  \end{center}

  \caption{The dependency of the disc-wind density and velocity on the wind
    acceleration parameter $\beta$. The wind density $\rho$ (right panel)
    and the poloidal velocity component $v_{p}$ (left panel) along the
    stream line starting from the mid point of the wind launching zone,
    i.e. $\left ( w,z \right ) = \left ( w_{\mathrm{mid}}, 0 \right )$
    where $w_{\mathrm{mid}} = \left( R_{\mathrm{wi}}+R_{\mathrm{wo}}
    \right ) / 2$, are shown as a function of the distance ($l$) from the
    wind launching point (c.f.~
    equations~\ref{eq:discwind-poloidal-velocity} and
    \ref{eq:discwind-density}). The azimuthal velocity component
    ($v_{\phi}$), which is independent of $\beta$
    (c.f. equation~\ref{eq:discwind-toroidal-velocity}), is also shown in
    the left panel for comparison. The density is normalised to the
    density $\rho_{0}$ at the wind launching point for the $\beta = 1.0$ case.
    The $v_{p}$ reaches the terminal velocity by 100~au for all
    $\beta$. In the far field ($l>10$~au), the density is approximately
    proportional to $\sim l^{-2}$. }
  
  \label{fig:discwind_vr_rho}

\end{figure*}

Assuming mass-flux conservation and using the velocity field defined
above, the disc wind density as a function of $w_{i}$ and $l$ can be
written as
\begin{equation}
  \rho\left(w_{i},l\right) =
  \frac{\dot{m}\left(w_{i}\right)}{v_{p}\left(w_{i},l\right)\,\left|\cos\delta\right|}\,\left\{ 
  \frac{d}{D\left(w_{i},l\right)\,\cos\delta}\right\} ^{2}
  \label{eq:discwind-density}
\end{equation} 
where $D$ and $\delta$ are the distance from the source point ($S$) to
a point along the stream line and the angle between the stream line
and the disc normal, respectively. Fig.~\ref{fig:discwind_vr_rho}
shows the density and the velocity components along the mid stream
line, passing through
$w_{i}=\left(R_{\mathrm{wi}}+R_{\mathrm{wo}}\right)/2$, on the disc
plane ($z=0$) for different values of the wind acceleration parameter
$\beta$.

\subsection{The accretion disc}

\label{sub:Accretion-disc}

We adopt a simple analytical accretion disc model, the $\alpha$-disc
`standard model' (\citealt{shakura:1973}; \citealt*{frank:2002}) with
the inner radius fixed at the inner radius of the magnetosphere at
equatorial plane. This corresponds to Region 4 in
Fig.~\ref{fig:config_discwind}.  

\subsubsection{Density and velocity}

The disc density distribution is given by 
\begin{equation}
  \rho_{\mathrm{d}}\left(w,z\right) = 
  \Sigma\left(w\right)\,\frac{1}{\sqrt{2\pi}h\left(w\right)}e^{-\left(\frac{z}{2h\left(w\right)}\right)^{2}}
  \label{eq:disc-density-function}
\end{equation}
where $w=\sqrt{x^{2}+y^{2}}$, $h$, $z$ and $\Sigma$ are the distance
from the symmetry axis, the scale height, the distance from the disc
plane, and the surface density at the mid-plane, respectively. The
mid-plane surface density and the scale height are given as:
\begin{equation}
  \Sigma\left(w\right) = 
  \frac{5M_{\mathrm{d}}}{8\pi R_{\mathrm{do}}^{2}}\, w^{-3/4}
  \label{eq:density-midplane}
\end{equation} 
where $R_{\mathrm{do}}$ and $M_{\mathrm{d}}$ are the disc radius
and the disc mass respectively, and
\begin{equation}
  h\left(w\right)=0.05\, R_{\mathrm{do}}\, w^{9/8}\,.
  \label{eq:scale-height}
\end{equation}
With these parameters, the disc is slightly flared. The inner radius
of the disc is set to $R_{\mathrm{di}}=R_{\mathrm{mi}}$ . The disc
mass, $M_{\mathrm{d}}$, is assumed to be 1/100 of the stellar mass
($M_{*}$), and the outer disc radius ($R_{\mathrm{do}}$) is
100~au. The velocity of the gas/dust in the disc is assumed to be
Keplerian.

Because of the geometrical constraints imposed by the magnetosphere
and the disc wind, the finite height of the $\alpha$ disc causes an
undesirable interface problem at the boundary; hence, the inner part
($R_{\mathrm{di}} < w < R_{\mathrm{wo}}$) of the disc is replaced by
the geometrically thin (but opaque) disc. Since the emissivity of the
disc at optical wavelengths is negligible, and the disc scale height
is relatively small in the inner part of disc, this simplification is
reasonable:  the most important factor in determining the shape of the 
H$\alpha$ profiles is the finite height of the disc at large radii for large
inclinations.

\subsubsection{Dust model}
\label{sub:dust-model}

In order to calculate the dust scattering and absorption cross section
as a function of wavelength, the optical constants of
\citet{draine:1984} for amorphous carbon grains and
\citet{hanner:1988} for silicate grains are used. The model uses the
{}``large grain'' dust model of \citet{wood:2002} in which the dust
grain size distribution is described by the following function:
\begin{equation}
  n\left(a\right)da = \left(C_{\mathrm{C}}+C_{\mathrm{Si}}\right)\,
  a^{-p}\exp\left[-\left(\frac{a}{a_{c}}\right)^{q}\right]da
  \label{eq:grain-dist-function}
\end{equation}
where $a$ is the grain size restricted between $a_{\mathrm{min}}$ and
$a_{\mathrm{max}}$, and $C_{\mathrm{C}}$ and $C_{\mathrm{Si}}$ are the
terms set by requiring the grains to completely deplete a solar
abundance of carbon and silicon. The parameters adopted in our model are:
$C_{\mathrm{C}}=1.32\times10^{-17}$,
$C_{\mathrm{Si}}=1.05\times10^{-17}$, $p=3.0$, $q=0.6$,
$a_{\mathrm{\mathrm{min}}}=0.1\,\mathrm{\mathrm{\mathrm{\mu m}}}$,
$a_{\mathrm{\mathrm{max}}}=1000\,\mathrm{\mathrm{\mathrm{\mu m}}}$,
and $a_{c}=50\,\mathrm{\mathrm{\mathrm{\mu m}}}$. This corresponds to
Model~1 of the dust model used by \citet{wood:2002}. See also their
Fig.~3.  The relative numbers of each grain is assumed to be that of
solar abundance, C/H$\sim 3.5\times10^{-4}$ \citep{anders:1989} and
Si/H$\sim 3.6\times10^{-5}$ \citep{grevesse:1993} which are similar to
values found in the ISM model of \citet*{mathis:1977} and
\citet*{kim:1994}.

\section{The radiative transfer model}

\label{sec:radiative-transfer-model}

We have extended the {\sc TORUS} radiative transfer code
(\citealt{harries:2000}; \citealt{kurosawa:2004a};
\citealt{symington:2005}) to include the multiple circumstellar
components described above. In previous calculations
\citep{symington:2005}, the model was used with a three-dimensional
(3-D) adaptive mesh refinement (AMR) grid to investigate the line
variability associated with rotational modulation of complex
geometrical configurations of magnetospheric inflow (see also
\citealt*{kurosawa:2005a}). We modified the code to handle a
two-dimensional (2-D) density distribution, and restricted our models
to be axi-symmetric.  Note that the velocity field is still in 3-D --
the third component can be calculated by using symmetry for a given
value of azimuthal angle.

The computation of H$\alpha$ is divided in two parts: (1)~the source
function calculation ($S_{\nu}$) and (2)~the observed flux/profile
calculation. In the first process, we have used the method of
\citet{klein:1978} (see also \citealt{rybicki:1978};
\citealt{hartmann:1994}) in which the Sobolev approximation method is
applied. The population of the bound states of hydrogen are assumed to
be in statistical equilibrium, and the gas to be in radiative
equilibrium. Our hydrogen atom model consist of 14 bound states and a
continuum. Readers are refer to \citet{harries:2000} for details.

Monte Carlo radiative transfer (e.g. \citealt{hillier:1991}), under
the Sobolev approximation, can be used when (1)~a large velocity
gradient is present in the gas flow, and (2)~the intrinsic line width
is negligible compared to the Doppler broadening of the line. In our
earlier models (\citealt{harries:2000}; \citealt{symington:2005}),
this method was adopted since these conditions are satisfied.
However, as noted and demonstrated by \citet{muzerolle:2001}, even
with a moderate mass-accretion rate ($~10^{-7}\,\mathrm{M_{\sun}\,
yr^{-1}}$), Stark broadening becomes important in the optically thick
H$\alpha$ line. \citet{muzerolle:2001} also pointed out that the
observed H$\alpha$ profiles from CTTS typically have the wings
extending to $500\,\mathrm{km\, s^{-1}}$(e.g.~\citealt{edwards:1994};
\citealt{reipurth:1996}) which cannot be explained by the infall
velocity alone.

We have implemented the broadening mechanism following the formalism
described by \citet{muzerolle:2001}. First, the emission and
absorption profiles are replaced from the Doppler to the Voigt
profile, which is defined as:
\begin{equation}
  H\left(a,y\right) \equiv
  \frac{a}{\pi}\int_{-\infty}^{\infty}\frac{e^{-y'^{2}}}{\left(y-y'\right)^{2}+a^{2}}\, dy'  
  \label{eq:voigt_profile_def}
\end{equation}
where $a=\Gamma/4\pi\Delta\nu_{\mathrm{D}}$,
$y=\left(\nu-\nu_{0}\right)/\Delta\nu_{\mathrm{D}}$, and
$y'=\left(\nu'-\nu_{0}\right)/\Delta\nu_{\mathrm{D}}$
(c.f. \citealt{mihalas:1978}).  $\nu_{0}$ is the line centre
frequency, and $\Delta\nu_{\mathrm{D}}$ is the Doppler line width of
hydrogen atom (due to its thermal motion) which is given by
$\Delta\nu_{\mathrm{D}}=\left(2kT/m_{\mathrm{H}}\right)^{1/2}\times\left(\nu_{0}/c\right)$
where $m_{\mathrm{H}}$ is the mass of a hydrogen atom. The damping
constant $\Gamma$, which depends on the physical condition of the gas,
is parameterised by \citet*{vernazza:1973} as follows:
\begin{eqnarray}
  \Gamma & = & C_{\mathrm{rad}} + C_{\mathrm{vdW}}
  \left(\frac{n_{\mathrm{HI}}}{10^{16}\,\mathrm{cm^{-3}}}\right)
  \left(\frac{T}{5000\,\mathrm{K}}\right)^{0.3}\nonumber \\
  &  & + \,
  C_{\mathrm{Stark}}\left(\frac{n_{e}}{10^{12}\,\mathrm{cm^{-3}}}\right)^{2/3}
  \label{eq:dampimg_constant_def}
\end{eqnarray}
where $n_{\mathrm{H\, I}}$ and $n_{e}$ are the number density of
neutral hydrogen atoms and that of free electrons. Also,
$C_{\mathrm{rad}}$, $C_{\mathrm{vdW}}$ and $C_{\mathrm{Stark}}$ are
natural broadening, van der Waals broadening, and linear Stark
broadening constants respectively.  We adopt this
parameterisation along with the values of broadening constants for
H$\alpha$ from \citet{luttermoser:1992},
i.e. $C_{\mathrm{rad}}=6.5\times10^{-4}$~\AA,
$C_{\mathrm{vdW}}=4.4\times10^{-4}$~\AA~ and
$C_{\mathrm{Stark}}=1.17\times10^{-3}$~\AA.  In terms of level
populations and the Voigt profile, the line opacity for the transition
$i\rightarrow j$ can be written as: 
\begin{equation}
  \chi_{l} = \frac{\pi^{1/2}e^{2}}{m_{e}c}f_{ij}n_{j}
  \left(1-\frac{g_{j}n_{i}}{g_{i}n_{j}}\right)H\left(a,y\right)
  \label{eq:line_opacity}
\end{equation}
where $f_{ij}$, $n_{i}$, $n_{j}$, $g_{i}$and $g_{j}$ are the
oscillator strength, the population of $i$-th level, the population of
$j$-th level, the degeneracy of the $i$-th level, and the degeneracy
of the $j$-th level respectively. $m_{e}$ and $e$ are the electron
mass and charge (c.f. \citealt{mihalas:1978}).

We further modified {\sc torus} by replacing the Monte Carlo line
transfer algorithm with a direct integration method
(c.f. \citealt{mihalas:1978}) for computing the observed flux as a
function of frequency.  The integration of the flux is performed in
the cylindrical coordinate system $\left(p,\, q,\, t\right)$ which is
obtained by rotating the original stellar coordinate system
$\left(\rho,\,\phi,\, z\right)$ around the $y$ axis by the inclination
of the line of sight.  Note that the $t$-axis coincides with the line
of sight with this rotation. The observed flux ($F_{\nu}$) is given
by:
\begin{equation}
  F_{\nu}=\frac{1}{4\pi d^{2}}
  \int_{0}^{p_{\mathrm{max}}}\int_{0}^{2\pi}p\,\sin q\,
  I_{\nu}\mathrm{\, d}q\,\mathrm{d}p
  \label{eq:flux_integral}
\end{equation}
where $d$, $p_{\mathrm{max}}$, and $I_{\nu}$ are the distance to
an observer, the maximum extent to the model space in the projected
(rotated) plane, and the specific intensity ($I_{\nu}$) in the direction
on observer at the outer boundary. For a given ray along $t$, the
specific intensity is given by:
\begin{equation}
       I_{\nu} = I_{0}e^{-\tau_{\infty}} + 
           \int_{t_{0}}^{t_{\infty}}S_{\nu}\left(t\right)\,
                e^{-\tau}\mathrm{d}\tau
 \label{eq:formal_sol_integral}
\end{equation}
where $I_{0}$ and $S_{\nu}$ are the intensity at the boundary on the
opposite to the observer and the source function
($\eta_{\nu}/\chi_{\nu}$) of the stellar atmosphere/wind at a
frequency $\nu$. For a ray which intersects with the stellar core,
$I_{0}$ is computed from the stellar atmosphere model of
\citet{kurucz:1979} as described in
Section~\ref{sub:Continuum-Source}, and $I_{0}=0$ otherwise. If the
the ray intersects with the hot ring on the stellar surface created by
the accretion stream, we set
$I_{0}=B_{\nu}\left(T_{\mathrm{ring}}\right)$ where $B_{\nu}$ is the
Planck function and $T_{\mathrm{ring}}$ is the temperature of the hot
ring.  The initial position of each ray is assigned to be at the
centre of the surface element ($\mathrm{d}A=p\,\sin
q\,\mathrm{d}q\,\mathrm{d}p$).  The code execution time is
proportional to $n_{p}\, n_{q}\, n_{\nu}$ where $n_{p}$ and $n_{q}$
are the number of cylindrical radial and angular points for the flux
integration, and $n_{\nu}$ is the number of frequency points. In the
models presented in the following section, $n_{p}=180$, $n_{q}=100$,
and $n_{\nu}=101$ are used unless specified otherwise. A linearly
spaced radial grid is used for the area where the ray intersects with
magnetosphere, and a logarithmically spaced grid is used for the wind
and the accretion disc regions.

The optical depth $\tau$ in equation~\ref{eq:formal_sol_integral}
is defined as: 
\[
\tau\left(t\right)\equiv\int_{t}^{\tau_{\infty}}\chi_{\nu}\left(t'\right)\,\mathrm{d}t'
\]
where $\chi_{\nu}$ is the opacity of media the ray passes through.
$\tau_{\infty}$ is the total optical depth measured from the initial
ray point to the observer (or to the outer boundary closer to the
observer). Initially, the optical depth segments $\mathrm{d\tau}$
are computed at the intersections of a ray with the original AMR grid
in which the opacity and emissivity information are stored. For high
optical depth models, additional points are inserted between the original
points along the ray, and $\eta_{\nu}$ are $\chi_{\nu}$ values are
interpolated to those points to ensure $\mathrm{d}\tau<0.05$ for
the all ray segments.

For a point in the magnetosphere and the wind flows, the emissivity
and the opacity of the media are given as:
\begin{equation}
  \left\{ 
  \begin{array}{rcl}
    \eta_{\nu} & = & \eta_{\mathrm{c}}^{\mathrm{H}}+\eta_{l}^{\mathrm{H}}\\
    \chi_{\nu} & = & \chi_{c}^{\mathrm{H}}+\chi_{l}^{\mathrm{H}}
    +\sigma_{\mathrm{es}}
  \end{array}
  \right.
  \label{eq:emiss_opa_gas}
\end{equation}
where $\eta_{\mathrm{c}}^{\mathrm{H}}$ and $\eta_{l}^{\mathrm{H}}$ are
the continuum and line emissivity of
hydrogen. $\chi_{c}^{\mathrm{H}}$, $\chi_{l}^{\mathrm{H}}$, and
$\sigma_{\mathrm{es}}$ are the continuum, line opacity
(equation~\ref{eq:line_opacity}) of hydrogen, and the electron
scattering opacity. Similarly, for a point in the accretion disc,
\begin{equation}
  \left\{ 
  \begin{array}{rcl}
    \eta_{\nu} & = & 0\\
    \chi_{\nu} & = & \kappa_{\mathrm{abs}}^{\mathrm{dust}}
    +\kappa_{\mathrm{sca}}^{\mathrm{dust}}
  \end{array}
  \right.
  \label{eq:emiss_opa_dust}
\end{equation}
where $\kappa_{\mathrm{abs}}^{\mathrm{dust}}$ and
$\kappa_{\mathrm{sca}}^{\mathrm{dust}}$ are the dust absorption, and
scattering opacity which are calculated using the dust property
described in Section~\ref{sub:dust-model}.  We neglect the
emissivity from the disc: since the disc masses of CTTS are rather small
($\sim 0.01\,\mathrm{M_{\sun}}$) and low temperature ($\lesssim 
1600\,\mathrm{K}$), the continuum flux contribution at H$\alpha$
wavelength is expected to be negligible (e.g.~\citealt{chiang:1997}).

\section{Results}
\label{sec:Results}

\subsection{Magnetosphere only models}
\label{sub:model-acc}

\begin{figure}

  \begin{center}
    
    \includegraphics[%
      clip,
      scale=0.45]{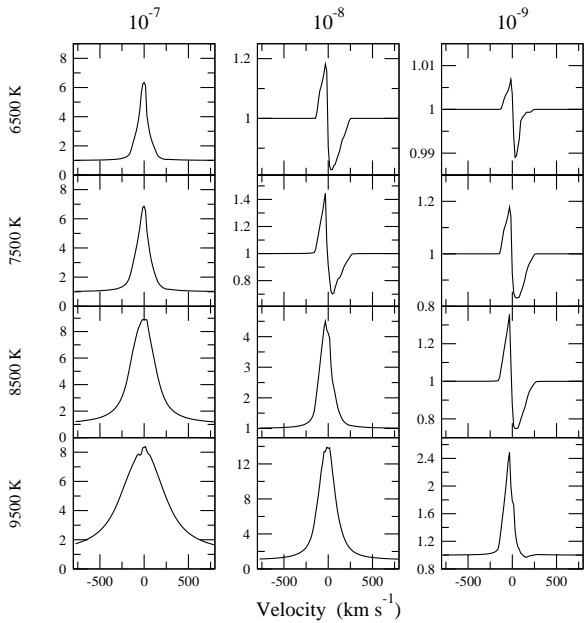} 
    
  \end{center}

  \caption{H$\alpha$ model profiles for wide ranges of mass accretion
    rate ($\dot{M}_\mathrm{acc}$) and temperature
    ($T_{\mathrm{max}}$). The profiles are computed using only the
    magnetospheric accretion flow (i.e. no outflow).  All the profiles
    are computed using the parameters of the reference model
    (Table~\ref{tab:std_parameters}) and inclination $i=55^{\circ}$. The
    temperature (indicated along the vertical axis) of the model increases
    from top to bottom, and the mass accretion rate (indicated by the
    values in $\mathrm{M_{\sun}\,yr^{-1}}$ along the top) decreases from
    left to right. }

  \label{fig:atlas_acc_models}

\end{figure}

\begin{figure}

  \begin{center}
    
    \includegraphics[%
      clip,
      scale=0.59]{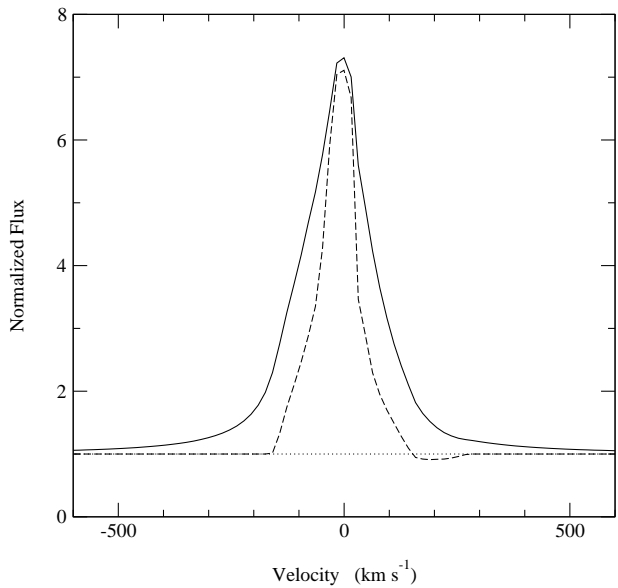}
    
  \end{center}
  
  \caption{The effect of line broadening on H$\alpha$. A model
    computed with a damping constant ($\Gamma$), described in
    Section~\ref{sec:radiative-transfer-model} (solid), is compared with
    a model with no damping, $\Gamma=0$ (dashed).  Both models are
    computed with $T_{\mathrm{max}}=7500~\Kelvin$, $i=55^{\circ}$, and the
    reference parameters given in Table~\ref{tab:std_parameters}. The two
    models have similar peak flux levels (around $V \sim 0~\kmps$), but
    the total flux and the EW of the line increased drastically for the
    model with the damping constant. The broad wings extend to $\sim \pm
    800\,\kmps$.  The redshifted absorption feature seen (very weakly) in
    the $\Gamma=0$ model is not seen in the model with the broadening.  }
  
  \label{fig:broadening}
  
\end{figure}

\begin{figure}

  \begin{center}
    
    \includegraphics[%
      clip,
      scale=0.59]{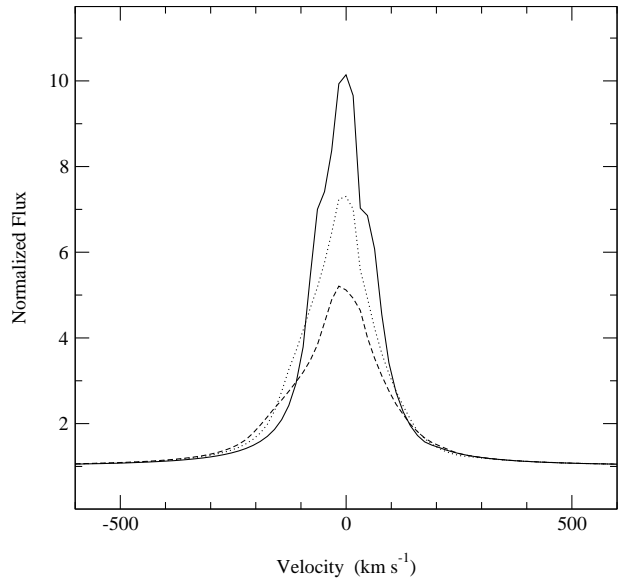}
    
  \end{center}

  \caption{The dependency of the H$\alpha$ profile on inclination ($i$).  The
    profiles are computed with a magnetospheric accretion flow only using
    the reference parameters given in Table~\ref{tab:std_parameters} and
    $T_{\mathrm{max}}=7500~\Kelvin$. The solid, dotted, and dashed lines
    are for $i=10^{\circ}$, $55^{\circ}$, and $80^{\circ}$
    respectively. As the inclination becomes larger, the peak flux and the
    equivalent width of the line become smaller.  }
  
  \label{fig:acc_inclination_effect}

\end{figure}

Using the reference model parameters (Table~\ref{tab:std_parameters})
for the central star and the magnetosphere, we examine the dependency
of H$\alpha$ on the temperature ($T_{\mathrm{max}}$) of the accretion
flow and the mass accretion rate ($\dot{M}_{\mathrm{acc}}$), as done
by \citet{muzerolle:2001} for H$\beta$. 
The temperature structure of \citet{hartmann:1994} is  adopted here,
and the parameter $T_{\mathrm{max}}$ sets the maximum temperature of the accretion
streams. 
We note that here we have
computed the hot ring temperature self-consistently, whereas
\citet{muzerolle:2001} used a constant hot ring temperature
($8000\,\mathrm{K}$) for most of their models.  The accretion
luminosity ($L_{\mathrm{acc}}$) for models with
$\dot{M}_{\mathrm{acc}}=10^{-7}\,\mathrm{M_{\sun}\, yr^{-1}}$ is about
a half of the photospheric luminosity.

H$\alpha$ line profiles for a range of $T_{\mathrm{max}}$ and
$\dot{M}_{\mathrm{acc}}$ are presented in  Fig.~\ref{fig:atlas_acc_models}. 
The  overall dependency on $T_{\mathrm{max}}$ and $\dot{M}_{\mathrm{acc}}$ is similar to that of
\citet{muzerolle:2001}. In general, the line strength weakens
as the accretion rate and the temperature drop.  The
red-shifted absorption becomes less visible for the higher accretion rate
and higher temperature models as it becomes filled-in by the stronger Stark-broadened
line-wing emission.

We demonstrate this effect in Fig.~\ref{fig:broadening}, which shows an example 
for an H$\alpha$  model with $T_{\mathrm{max}}=7500\,\mathrm{K}$
and $\dot{M}_{\mathrm{acc}}=10^{-7}\,\mathrm{M_{\sun}\, yr^{-1}}$, both with and without
damping.  Although the maximum flux of the model with broadening is almost identical to that of the
model with no damping constant ($\Gamma=0$), a significant increase of
the line flux in both red and blue wings is seen. The weak red-shifted
absorption component (which is a signature of the magnetospheric accretion)
is weakened or eliminated by the flux in broadened wing.

Table~\ref{tab:ha_ew_acc_models} shows the equivalent width (EW) for
the models in Fig.~\ref{fig:atlas_acc_models}.  The EWs for half the
models fall within the range of EWs ($\sim 3$ to $\sim 160$~\AA)
measured by \citet{alencar:2000}. The EWs of models 
with the lowest mass accretion rates fall below the minimum EW
observed by  \citet{alencar:2000}, while several models would
be designated as weak-lined T~Tauri stars (WTTS) using the traditional
10\,\AA\ cut-off \citep[e.g.][]{herbig:1988}.
\citet{white:2003} empirically showed that the full width of H$\alpha$ at 10
per~cent of its peak flux (10 per~cent width) is a better
indicator for an accretion than the EW criteria. They proposed that if
a T~Tauri star shows a 10 per~cent width greater than
$270\,\mathrm{km\,s^{-1}}$, the star should be a CTTS. The 10 per~cent
widths of the model H$\alpha$ profiles
(Fig.~\ref{fig:atlas_acc_models}) are summarised in
Table~\ref{tab:ha_ten_percent_acc_models}.  Using the criteria of
\citet{white:2003}, half of the models shown in the figure can be
classified as CTTS, and the other half as WTTS. However, the profiles corresponding
to the WTTS designations have an inverse P-Cygni morphology that is rarely
seen in observations (\citealt{reipurth:1996} and
Section~\ref{sub:Classification-scheme-proposed}).

The dependency of the line profile on inclination angle ($i$) is
demonstrated in Fig.~\ref{fig:acc_inclination_effect}. The model
has  $T_{\mathrm{max}}=7500\,\mathrm{K}$ and
$\dot{M}_{\mathrm{acc}}=10^{-7}\,\mathrm{M_{\sun}\, yr^{-1}}$.  The
figure shows that the peak (normalised) flux decreases as the
inclination angle increases, as does the EW.  
The line flux and the EWs decrease mainly because (1)~the fraction of
the accretion stream blocked by the photosphere increases as the
inclination increases (c.f.~Fig.~\ref{fig:config_discwind}, lower-left
part of the accretion stream i.e.~$z<0$ and $x<0$), and (2)~the
fraction of the accretion stream blocked by the inner part of the
accretion disc increases as the inclination increases
(c.f.~Fig.~\ref{fig:config_discwind}, lower-right part of accretion
stream i.e.~$z<0$ and $x>0$). While the photosphere mostly blocks the
blue-shifted part of the stream, the inner disc blocks the red-shifted
part.
The line EW in Fig.~\ref{fig:acc_inclination_effect} changes from
32 to 21\,\AA\, as the inclination changes from $10^\circ$ to
$80^\circ$, i.e.~a fractional change of 0.65.  On the other hand,
the 10~per~cent width of the line changes from 290 to
470~$\kmps$; hence, the fractional change is 1.6.  
A similar EW dependence on
inclination angle is found for the models with different 
magnetospheric temperatures and the mass-accretion rates.

Because of the geometry of the
magnetospheric accretion (c.f. Fig.~\ref{fig:config_discwind}) and of
the presence of the gas with the highest velocity close to the stellar
surface, the highest blue-shifted line-of-sight velocity is seen 
only at the high inclination angles. For example, the
accretion stream close to the photosphere ($\sim 1R_{*}$ above the
photosphere) on the upper-left
quadrant in Fig.~\ref{fig:config_discwind} ($x<0$ and $z>0$) will be
more aligned with the line-of-sight of an observer in the upper-right
quadrant in the same figure ($x>0$ and $z>0$) as the inclination
increases. This results in a slight increase in the blue wing flux for
a higher inclination model as seen in
Fig.~\ref{fig:acc_inclination_effect}.  

Our models show a blue-shifted line peak and
a blueward asymmetry caused by occultation of the accretion
flow by the stellar photosphere (see also \citealt{hartmann:1994};
\citealt{muzerolle:2001}). However, \citealt{alencar:2000} (see their fig.~9)
found a substantial fraction
of the observed H$\alpha$ profiles show a red-shifted
peak. Furthermore, a recent study by
\citet{appenzeller:2005} demonstrated  that the equivalent width of H$\alpha$
from CTTS {\em increases} as the inclination angle increases. 
Clearly a magnetospheric accretion only model cannot explain the
full properties of the H$\alpha$ profiles in CTTS.

\begin{table}

\caption{Summary of H$\alpha$ equivalent widths (\AA) from the
magnetospheric accretion flow models shown in
Fig.~\ref{fig:atlas_acc_models}. Note that the equivalent widths are
positive when the lines are in emission.}

\label{tab:ha_ew_acc_models}

\begin{center}

\begin{tabular}{cccc}
\hline 
$T_{\mathrm{max}}\,\,\left(\mathrm{K}\right)$&
&
$\dot{M}_{\mathrm{acc}}$ ~$\left(\mathrm{M_{\sun}\, yr^{-1}}\right)$&
\tabularnewline
&
$10^{-7}$&
$10^{-8}$&
$10^{-9}$\tabularnewline
\hline 
$6500$&
$17.9$&
$0.1$&
$0.0$\tabularnewline
$7500$&
$25.2$&
$-0.9$&
$-0.5$\tabularnewline
$8500$&
$68.3$&
$6.5$&
$-0.7$\tabularnewline
$9500$&
$98.6$&
$52.4$&
$1.3$\tabularnewline
\hline
\end{tabular}

\end{center}

\end{table}

\begin{table}

\caption{Summary of 10 per~cent widths ($\mathrm{km\,s^{-1}}$) from the
magnetospheric accretion flow models shown in
Fig.~\ref{fig:atlas_acc_models}.}

\label{tab:ha_ten_percent_acc_models}

\begin{center}

\begin{tabular}{cccc}
\hline 
$T_{\mathrm{max}}\,\,\left(\mathrm{K}\right)$&
&
$\dot{M}_{\mathrm{acc}}$ ~$\left(\mathrm{M_{\sun}\, yr^{-1}}\right)$&
\tabularnewline
&
$10^{-7}$&
$10^{-8}$&
$10^{-9}$\tabularnewline
\hline 
$6500$&
$300$&
$120$&
$100$\tabularnewline
$7500$&
$360$&
$145$&
$120$\tabularnewline
$8500$&
$790$&
$330$&
$130$\tabularnewline
$9500$&
$1530$&
$520$&
$200$\tabularnewline
\hline
\end{tabular}

\end{center}

\end{table}

\subsection{Disc-wind only models}
\label{sub:result-discwind-knigge}

\begin{figure*}

  \begin{center}
    
    \includegraphics[%
      clip,
      scale=0.9]{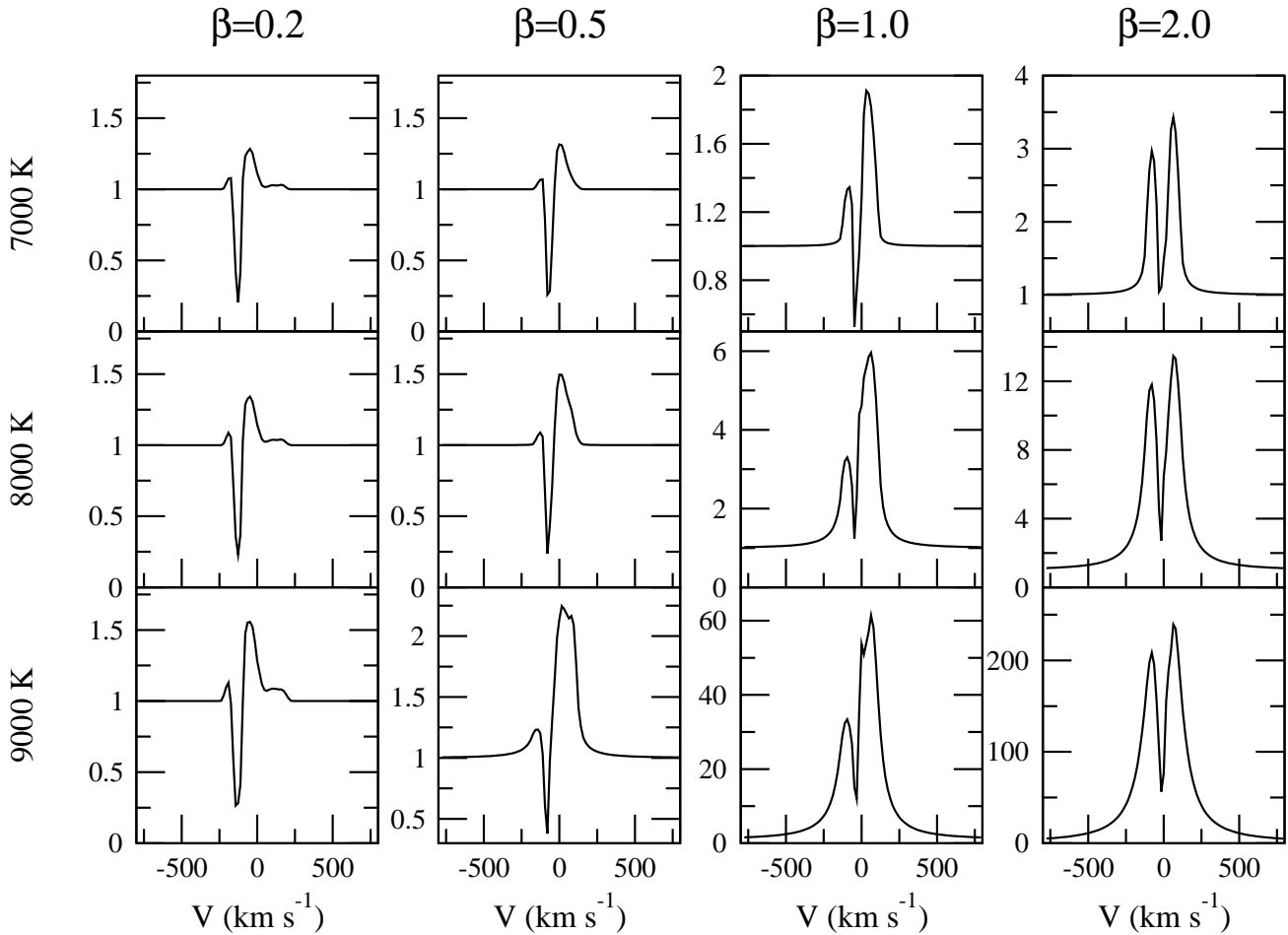}
    
  \end{center}

  \caption{H$\alpha$ profiles computed with the disc-wind only.
     The wind mass-loss rate is fixed at
    $\dot{M}_{\mathrm{wind}}=10^{-8}\,\mathrm{M_{\sun}\,yr^{-1}}$, and the line
    profiles are computed with different combinations of the wind
    acceleration rate ($\beta$) and isothermal disc-wind temperature
    ($T_{\mathrm{wind}}$). The wind emission grows as the values of
    $\beta$ and $T_{\mathrm{wind}}$ increase.   The position
    of the wind absorption moves toward the line centre as $\beta$ becomes
    larger.   }

  \label{fig:atlas_discwind_only}

\end{figure*}

In this section, we examine profiles produced using 
the disc wind outflow model described in
Section~\ref{sub:model-disc-wind}. 
The parameters used for the
central star are as in Section~\ref{sub:Continuum-Source}, and the
disc-wind parameters are summarised in
Table~\ref{tab:std_discwind_parameters}. Although the line is potentially
sensitive to the temperature structure of the wind, determination of a
self-consistent wind temperature is beyond the scope of this
paper. Readers are refereed to \citet{hartmann:1982} in which the wind
temperature structure is determined by balancing the radiative cooling
rate (assuming optically thin) with the MHD wave heating rate.
We pragmatically assume that the wind is isothermal at $T_{\mathrm{wind}}$.

Initially we examine the characteristics of the H$\alpha$ profile as a
function of the wind acceleration parameter $\beta$ and the isothermal
wind temperature $T_{\mathrm{wind}}$.  The mass-loss rate is kept
constant at
$\dot{M}_{\mathrm{wind}}=10^{-8}\,\mathrm{M_{\sun}\,yr^{-1}}$.
Fig.~\ref{fig:atlas_discwind_only} shows the model profiles computed
for the wind temperatures between $7000\,\mathrm{K}$ and
$9000\,\mathrm{K}$, and those with $\beta$ between $0.2$ and $2.0$,
computed at inclination $i=55^{\circ}$.
The morphology of the profile exhibited by the model changes from
FU-Ori type (c.f.~\citealt{reipurth:1996}) for the models with small
$\beta$ to the stronger emission types, which are more commonly seen in
the observed H$\alpha$ (c.f.~\citealt{reipurth:1996}), as the value of
$\beta$ increases.  
Classical P~Cygni profiles are prominent
in the models with lower values of $T_{\mathrm{wind}}$ and $\beta$
(a colder and faster accelerating wind). As the value of $\beta$
increases, the position of the blue-shifted absorption component moves
toward the line centre. The blue edge of the absorption component in
the fastest accelerating wind 
($\beta=0.2$) model is almost at the terminal velocity ($v_{\infty}\sim
160\,\kmps$) of the disc wind (c.f.~Fig.~\ref{fig:discwind_vr_rho}), that
for the slowest wind acceleration model 
($\beta=2.0$) is located close the line centre ($\sim 20\,\kmps$).

The intensity of the emission component increases as
$\beta$ increases for a fixed $T_{\mathrm{wind}}$. This is mainly
because the density near the photosphere increases as the value of
$\beta$ increases (Fig.~\ref{fig:discwind_vr_rho}); hence, the
emissivity increases. Most of the profiles show a red-ward asymmetry, mainly
due to the presence of the classical P-Cygni absorption. However we find
that the disc-wind models with the fastest acceleration ($\beta=0.2$ and $0.5$)
and viewed at low inclinations, show a strong blue asymmetry due to
occultation of the receding half of the outflow by the optically thick disc.

\begin{figure}

  \begin{center}

    \includegraphics[%
      clip,
      scale=0.45]{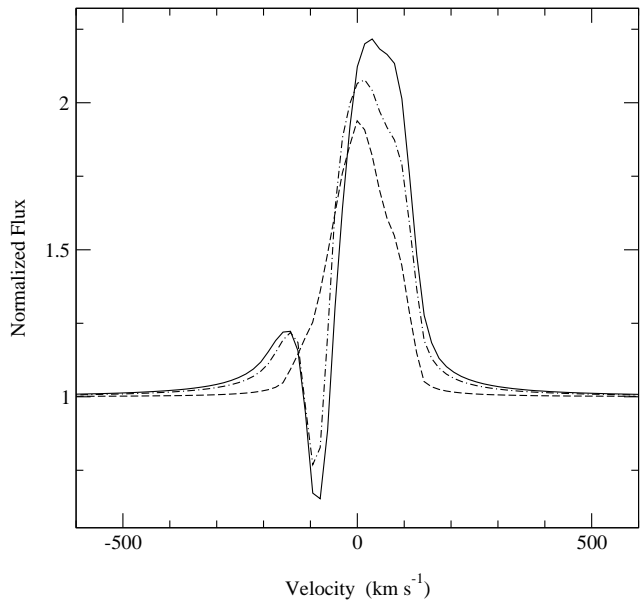}

  \end{center}

  \caption{The dependency of the H$\alpha$ profile on the source
    displacement parameter $d$. The line profiles are computed with 
    the wind parameter $T_{\mathrm{wind}} = 9000\Kelvin$ and $\beta  =
    0.5$ for $i=55^{\circ}$.  The profiles in the figure use $d = 
   2~R_{*}$ (dashed),   $7~R_{*}$ (dash-dot) and $22~R_{*}$
    (solid). All the other parameters are same as in
    Fig.~\ref{fig:atlas_discwind_only}.  As the wind becomes more
    collimated (larger $d$), the wind emission becomes stronger.  
    The P-Cygni absorption disappears for the model with a small
    collimation (e.g.~$d = 2~R_{*}$ case) since the opening angle of
    the disc wind is larger than the inclination angle. }
  
  \label{fig:discwind_d_effect}

\end{figure}

We now examine the sensitivity of the model profiles to the wind
collimation or the source displacement parameter $d$. As we can see
from Fig.~\ref{fig:config_discwind}, the disc wind
geometry/collimation is determined by $d$ and the wind launching radii
($R_{\mathrm{wi}}$ and $R_{\mathrm{wo}}$). Here, we fix the values of
$R_{\mathrm{wi}}$ and $R_{\mathrm{wo}}$ as in
Table~\ref{tab:std_discwind_parameters}, and inspect the model
dependency on $d$ alone.  The minimum wind opening angle (measured
from the symmetry axis) is given by
$\theta_{\mathrm{open}}=\arctan\left( R_{\mathrm{wi}}/d \right)$.
Fig.~\ref{fig:discwind_d_effect} shows the profiles computed for $d
= 2~R_{*}$, $7~R_{*}$ and $22~R_{*}$ for the inclination angle
$i=55^{\circ}$.  As the wind becomes more collimated (larger $d$), the
wind opening angle becomes smaller and the wind density is enhanced
towards the symmetry axis ($z$); hence, the wind emission increases.
When the collimation of the wind is small (as in the $d = 2~R_{*}$
case), the P-Cygni absorption is absent from the profile because the
opening angle of the disc wind is greater than the inclination angle.
For the disc wind models presented in
Fig.~\ref{fig:atlas_discwind_only}, we use $d=22~R_{*}$ 
in order to have a moderate amount of wind absorption at a moderate
inclination angle ($i\sim60^\circ$).

\citet{blandford:1982} showed that in order to launch a wind from a
disc plane magneto-centrifugally, the launching angle ($\theta_{L}$,
measured from the equatorial disc plane) must be less than $60^{\circ}$.
 In our models, the wind launching zone is restricted in
between $R_{\mathrm{wi}} < w_{i} < R_{\mathrm{wo}}$ (see
Fig.~\ref{fig:config_discwind} and Table~\ref{tab:std_parameters}) with
$R_{\mathrm{wi}} = 3\,R_{*} $ and 
$R_{\mathrm{wo}}=1$~au.  This sets the upper limit of the source displacement
distance ($d$) to be around $5.2\,R_{*}$ if we want to satisfy the
launching angle condition of \citet{blandford:1982} for all the
stream lines in our disc wind model.  In the models presented in
Fig.~\ref{fig:atlas_discwind_only} (and later in
Fig.~\ref{fig:atlas_hybrid} in next section), we have adopted 
$d =22\, R_{*}$ instead. This means that some part of the disc
wind ($3\,R_{*} < w_{i} < 12.7\,R_{*}$) may not be able to be launched
from the disc plane according to the model of \citet{blandford:1982}.
We have used a larger launching angle or a larger source displacement
distance for the following two reasons.

First, we want to simulate a more \emph{collimated} wind (toward
the symmetry axis).  Since the stream lines in our simple disc
wind are represented by straight lines, the wind never bends
toward the symmetry axis ($z$-axis) as the distance from the
disc increases unlike the model of \citet{blandford:1982}. We
therefore used a larger launching angle to simulate 
the stream lines in the far field (at a larger distance from
the disc plane) in their model.  This is an intrinsic problem
with approximating the stream lines with straight lines.

Second, the larger source displacement model would provide a greater
consistency with the fraction of H$\alpha$ profiles affected by wind
absorption seen in observations.  According to the observations of
\citet{reipurth:1996} and \citet{alencar:2000}, more than 50 per cent
of CTTS show wind absorption features in their H$\alpha$
profiles. In our model, if the launching angle is too small
($\theta_{\mathrm{L}}<60^{\circ}$), the wind absorption feature would not be seen or
would be very small for profiles computed with $i<60^{\circ}$, as
demonstrated in Fig.~\ref{fig:discwind_d_effect}.  The situation will
however change if the field/stream line is not straight and bending
towards the symmetry axis as the wind travels higher above the disc.

In addition to the models given in Figs.~\ref{fig:atlas_discwind_only}
and \ref{fig:atlas_hybrid}, we have computed the models with $d = 5\,
R_{*}$ (not shown here) which satisfy the wind launching condition
of \citet{blandford:1982} in all parts of the disc wind zone in our
model.  These models showed that the wind absorption and emission
features weaken, but overall morphology dependence of the profiles on
the wind temperature and the wind acceleration rate is similar to
what we found in the models shown in
Figs.~\ref{fig:atlas_discwind_only} and \ref{fig:atlas_hybrid}.  The
minimum inclination angle at which the wind absorption becomes
important is larger than that of the models with a larger source
displacement ($d = 22\, R_{*}$) as expected.

\begin{figure}

  \begin{center}
    
    \includegraphics[%
      clip,
      scale=0.45]{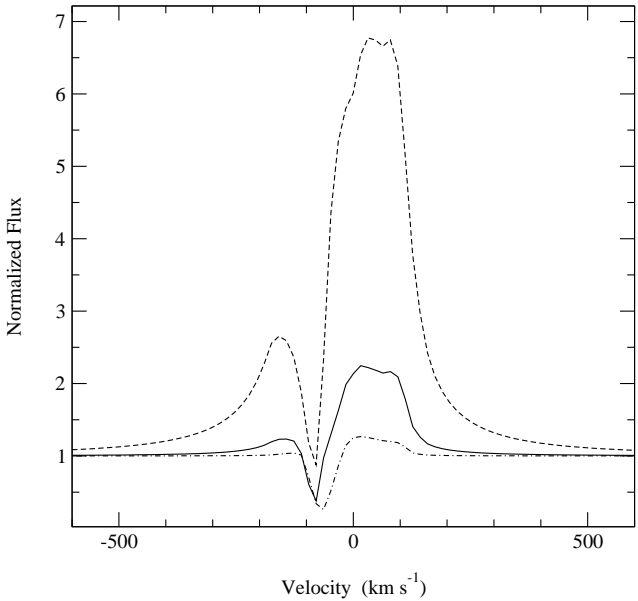}
    
  \end{center}
  
  \caption{The effect of the wind mass-loss rate.  The profiles computed
    with $\left(T_{\mathrm{wind}},~\beta \right) = \left( 9000\Kelvin,
    0.5\right)$, and $\dot{M}_{\mathrm{wind}} = 
    2.0 \times 10^{-8}~\mathrm{M_{\sun}\,yr^{-1}}$\, (dashed), 
   $1.0 \times 10^{-8}~\mathrm{M_{\sun}\,yr^{-1}}$\, (solid), and 
   $0.5 \times 10^{-8}~\mathrm{M_{\sun}\,yr^{-1}}$\, (dash-dot) 
    are shown above. With all other parameter fixed, 
    the mass-loss rate behaves as a scaling factor for the wind
    density (c.f.~equation~\ref{eq:discwind-density}).}
  
  \label{fig:discwind_mdot_effect}

\end{figure}


Finally, the effect of varying the wind mass-loss rate is shown in
Fig.~\ref{fig:discwind_mdot_effect}. With all other parameters fixed,
$\dot{M}_{\mathrm{wind}}$ acts as a scaling factor for the wind
density (equation~\ref{eq:discwind-density}). The most important process
that populates the $n=3$ level is recombination, and therefore the
line emissivity varies as the square of the density (assuming that the
ionisation fraction remains constant); hence the line is very sensitive to 
changes in $\dot{M}_{\mathrm{wind}}$.  We indeed find that  the line flux of
the models is approximately proportional to the square
of the mass-loss rate.

\begin{table}

\caption{A summary of the reference disc-wind model parameters used
in Sections~\ref{sub:result-discwind-knigge} and \ref{sub:result-discwind-hybrid}.
See also Section~\ref{sub:model-disc-wind}.}

\label{tab:std_discwind_parameters}

\begin{center}

\begin{tabular}{cccccc}
\hline 
$R_{\mathrm{wi}}$&
$R_{\mathrm{wo}}$&
$\dot{M}_{\mathrm{wind}}$&
$p$&
$f$&
$R_{\mathrm{s}}$\tabularnewline
$\left[R_{*}\right]$&
$\left[\mathrm{au}\right]$&
$\left[\mathrm{M_{\sun}\,yr^{-1}}\right]$&
$[-]$&
$[-]$&
$\left[R_{*}\right]$\tabularnewline
\hline
$3.0$&
$1.0$&
$10^{-8}$&
$-7/2$&
$2.0$&
$30$\tabularnewline
\hline
\end{tabular}

\end{center}

\end{table}

\subsection{Disc-wind-magnetosphere hybrid models}
\label{sub:result-discwind-hybrid}

\begin{figure*}

  \begin{center}
    
    \includegraphics[%
      clip,
      scale=0.9]{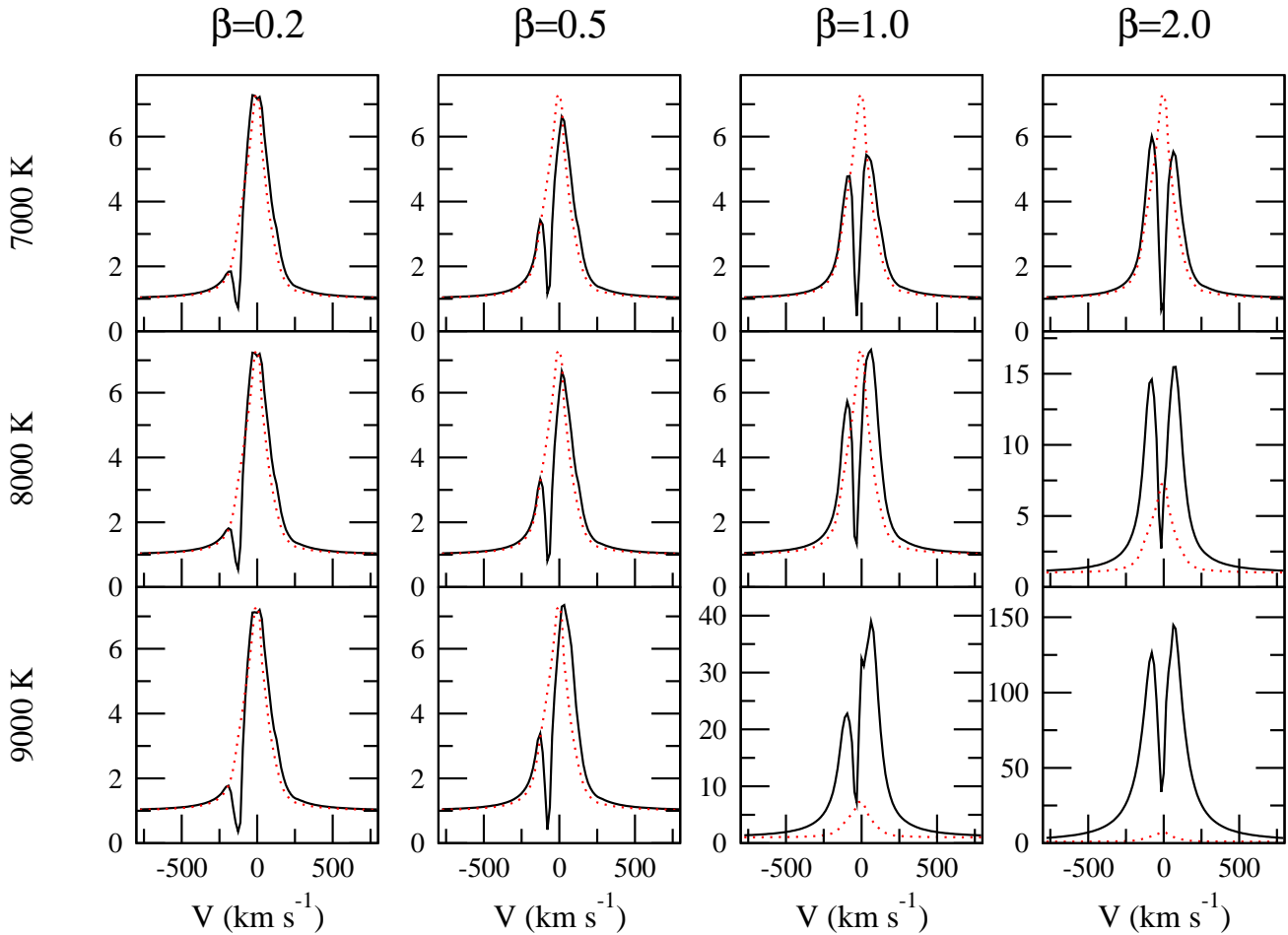}
    
  \end{center}

  \caption{The disc-wind-magnetosphere hybrid model. These models
  (solid lines) have the  same parameters as
    Fig.~\ref{fig:atlas_discwind_only}, but  also
    include the magnetospheric accretion flow
    ($T_{\mathrm{\max}}=7500~\mathrm{K}$ and
    $\dot{M}_{\mathrm{acc}}=10^{-7}\,\mathrm{M_{\sun}\,yr^{-1}}$). While
    the emission from the magnetosphere dominates for the models with
    smaller $\beta$ and $T_{\mathrm{wind}}$, the emission from the wind dominates the
    profiles for models with larger $\beta$ and $T_{\mathrm{wind}}$.  The contribution
    of the magnetosphere can be judged from the profiles computed with
    the magnetosphere only (dotted lines).}

  \label{fig:atlas_hybrid}

\end{figure*}

We now consider models computed with a combination of the
disc-wind and magnetospheric accretion components. The parameters
used for the magnetosphere are  same as in 
Fig.~\ref{fig:acc_inclination_effect} (with $i=55^{\circ}$ ), and  the mass-loss rate of
the disc wind is 
$\dot{M}_{\mathrm{wind}}=10^{-8}\,\mathrm{M_{\sun}\, yr^{-1}}$
(i.e.~$\mu=\dot{M}_{\mathrm{wind}}/\dot{M}_{\mathrm{acc}}=0.1$).
Fig.~\ref{fig:atlas_hybrid} shows the models profiles
computed (at $i=55^\circ$) using the same ranges of the disc-wind temperature and
wind acceleration parameter as in the previous cases.

As in the disc-wind only models
(Fig.~\ref{fig:atlas_discwind_only}), the location of the absorption
component moves toward the line centre as the value of $\beta$
increases.  The figure shows that in the models with smaller $\beta$
and $T_{\mathrm{wind}}$, the line emission from the magnetosphere
dominates, while this situation reverses for the models with larger
$\beta$ and $T_{\mathrm{wind}}$. The model with
$\left(T_{\mathrm{wind}},\,\beta\right)=\left(9000\,\mathrm{K},\,2.0\right)$
results in an H$\alpha$ profile that is far too strong to be
compatible with observations (\citealt{reipurth:1996}; 
\citealt{alencar:2000}); however, the lower mass-loss rate models
(while keeping $\mu$ constant) produce profiles with the line
strengths comparable to observations (c.f.~Appendix~\ref{sec:appendix}).
Although the models in Fig.~\ref{fig:atlas_hybrid} are computed with
limited ranges of $T_{\mathrm{wind}}$, $\beta$ and a fixed $i$, the
resulting profiles exhibit a wide variety of line profile shapes, many
of which are similar to the types of H$\alpha$ profiles seen in the
observations (c.f.~\citealt{reipurth:1996}).

In order to quantify the relative contributions of the wind emission and
 the magnetosphere, we have computed the ratio of the EW of the
hybrid model in Fig.~\ref{fig:atlas_hybrid} to the EW of the
magnetosphere only model (Table~\ref{tab:ew_ratio}).
This  ratio falls below 1.0 when the wind only contributes
to the line absorption, but not to the emission. 
In some models (e.g.~with $T_{\mathrm{wind}}=8000\,\Kelvin$ and
$\beta=2.0$), the contribution of the wind to the EW is much larger
than that of the magnetosphere (EW ratio is 3.2). This clearly
demonstrates that the difficulty of using the EW of H$\alpha$ alone as
an accretion measure.   We have also seen in Fig.~\ref{fig:atlas_discwind_only}
that disc wind models {\em without}  a magnetosphere can produce an
H$\alpha$ line with significantly large EW.

\begin{figure}

  \begin{center}

    \includegraphics[%
      clip,
      scale=0.45]{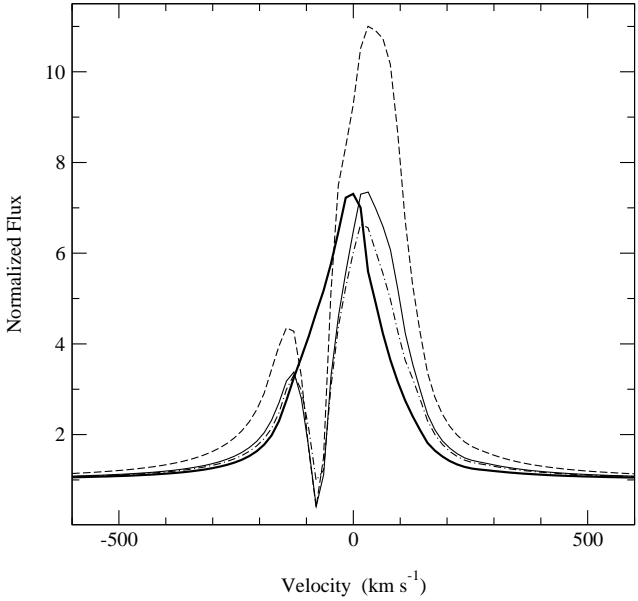}
    
  \end{center}

  \caption{The effect of the mass-loss rate to mass-accretion rate
    ratio ($\mu=\dot{M}_{\mathrm{wind}}/\dot{M}_{\mathrm{acc}}$).  The
    H$\alpha$ profile computed with only the magnetospheric accretion
    (thick solid) is compared to the disc-wind-magnetosphere
    hybrid models for $\mu =$ 0.05 (dash-dot), 0.1
    (solid) and 0.2 (dash). The magnetosphere used here has
    $T_{\mathrm{max}}=7500\,\Kelvin$ and
    $\dot{M}_{\mathrm{acc}}=10^{-7}\,\mathrm{M_{\sun}\,yr^{-1}}$.  The
    temperature and the acceleration parameter of the disc wind are
    $T_{\mathrm{wind}}=9000\,\Kelvin$ and $\beta=0.5$ respectively.
    All the models are computed with the inclination angle
    $i=55^{\circ}$.  As $\mu$ increases the emission component becomes
    stronger and the wings becomes broader, as expected. The
    absorption component is relatively insensitive to the value of
    $\mu$ in this example. The wide range of $\mu$ reasonably
    reproduces the profiles similar to ones seen in observations
    (e.g.~\citealt{reipurth:1996}).}

  \label{fig:discwind_mu_effect}

\end{figure}

An examination of the profiles in
Figs.~\ref{fig:atlas_discwind_only} and \ref{fig:atlas_hybrid} reveals
some problems 
with uniqueness that must inevitably occur when trying to assign an
individual profile as wind or accretion dominated i.e. very similar
profiles may arise from very different circumstellar geometries. For
example, the disc-wind only model with
$\left(T_{\mathrm{wind}},\,\beta\right)=\left(8000\,\mathrm{K},\,1.0\right)$
in Fig.~\ref{fig:atlas_discwind_only} and the accretion dominated
model with
$\left(T_{\mathrm{wind}},\,\beta\right)=\left(9000\,\mathrm{K},\,0.5\right)$
in Fig.~\ref{fig:atlas_hybrid} give very similar
profiles.

The effect of changing the ratio of the wind mass-loss rate to
mass-accretion rate
($\mu=\dot{M}_{\mathrm{wind}}/\dot{M}_{\mathrm{acc}}$) is demonstrated
in Fig.~\ref{fig:discwind_mu_effect}. With a fixed value of
mass-accretion rate
($\dot{M}_{\mathrm{acc}}=10^{-7}\,\mathrm{M_{\sun}\, yr^{-1}}$), the
mass-loss rate is varied. The figure shows that as $\mu$ increases the
P-Cygni absorption deepens only slightly, and the position of minimum
flux in the absorption trough appears to remain at same location.  On the other
hand, the flux peak increases and the line wings become stronger as
$\mu$ increases.  Although MHD models suggest
(e.g.~\citealt{koenigl:2000}) that $\mu \sim 0.1$, the figure suggest
that models with a relatively wide range of $\mu$ (e.g.~between 0.05 and
0.2 in this case) produce profiles similar to observations.

\begin{figure}

  \begin{center}

    \includegraphics[%
      clip,
      scale=0.45,angle=270]{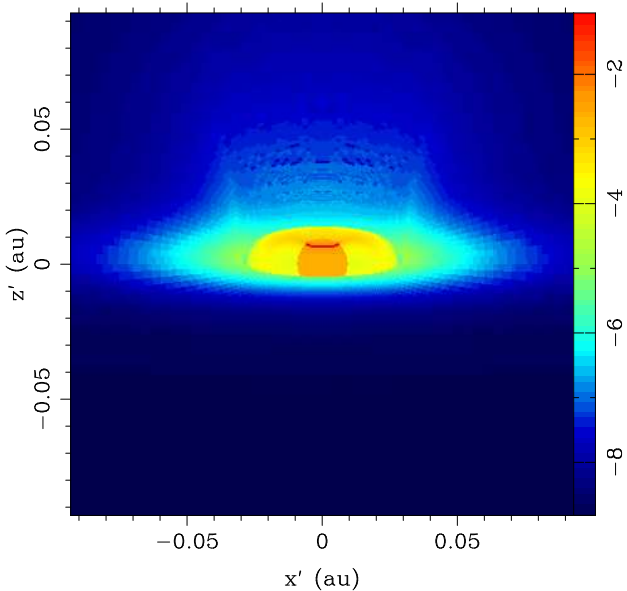}
    
  \end{center}

  \caption{An H$\alpha$ image of the disc-wind hybrid model viewed at an inclination
  of 80$^\circ$. The false-colour scale shows logarithmic (base 10)
  surface brightness (arbitrary units) and the axes are linear
  dimensions in au. The parameters used for the magnetosphere are as
  the reference model in Table~\ref{tab:std_parameters}. The wind
  temperature of $T_{\mathrm{max}}=7000~\Kelvin$ and the wind
  acceleration parameter $\beta=1.0$ are adopted here.  } 

  \label{fig:discwind_hybrid_ha_image}

\end{figure}

By inspecting the model H$\alpha$ images computed
(Fig.~\ref{fig:discwind_hybrid_ha_image}), we have 
found that the extent of the H$\alpha$ emission region is relatively
small (i.e.~$h\sim 0.1\,\mathrm{au}$ or $10\, R_{*}$ where $h$ is the
vertical distance from the disc) when compared to the
spectro-astrometric observations of \citet{takami:2003} 
for RU~Lup and CS Cha which show 1--5~au scale outflows%
\footnote{Their observations also show some objects
(e.g~Z CMa and AS~353A) displaying the outflows in larger scale
($>50\,\mathrm{au}$); however, this could be formed in shocks rather
than MHD-wave heating (e.g. \citealt{hartmann:1982}) or X-ray heating
(e.g.~\citealt{shang:2002}). %
}.  In our model, the wind emission flux drops by $\sim10^{4}$ by
the time the wind reaches $h \approx 0.1\,\mathrm{au}$. The extent of the line
emission region becomes slightly larger ($\sim0.4$~au) if a slower
wind acceleration rate (e,g,~$\beta=4$) is used. The discrepancy
between the observations and the model is perhaps caused by the lack
of the material connecting the \emph{disc wind} to a collimated
\emph{jet} in our model, which may be excited by alternative heating
mechanism (e.g.~shocks) thus producing observable H$\alpha$ emission
at large distances from the central object.

\begin{table}

  \caption{The ratio of the line equivalent width (EW) of the
    disc-wind magnetosphere hybrid models in
    Fig.~\ref{fig:atlas_hybrid} to the EW of the magnetosphere only model. } 

\label{tab:ew_ratio}

\begin{center}

\begin{tabular}{lccccc}
\hline 
$T_{\mathrm{max}}\,\,\left(\mathrm{K}\right)$&
&
$\beta$&
&
&
\tabularnewline
&
&
$0.2$&
$0.5$&
$1.0$&
$2.0$\tabularnewline
\hline 
$7000$&
&
$1.1$&
$.94$&
$.89$&
$.96$\tabularnewline
$8000$&
&
$1.1$&
$.92$&
$1.2$&
$3.2$\tabularnewline
$9000$&
&
$1.0$&
$1.1$&
$7.5$&
$36.$\tabularnewline
\hline
\end{tabular}

\end{center}

\end{table}

\section{Discussion}

\label{sec:Discussion}

\subsection{The morphological classification
 scheme proposed by Reipurth et. al (1996)}

\label{sub:Classification-scheme-proposed}

\begin{figure}

  \begin{center}

    \includegraphics[%
      clip,
      scale=0.65]{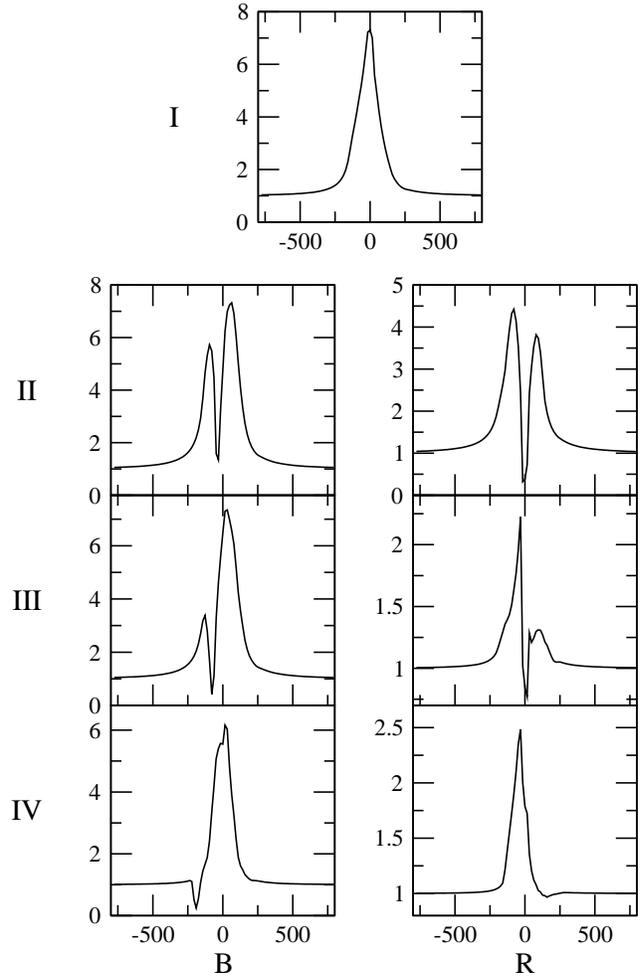}
    
  \end{center}
  
  \caption{Sample H$\alpha$ model profiles which characterise the
    classification scheme of \citet{reipurth:1996}. The combination of
    magnetospheric accretion flow, the accretion disc, and the disc-wind
    can reproduce the wide range of H$\alpha$ profiles seen
    in observations.  The model parameters used for each profile along with
    short comments are given in
    Table~\ref{tab:classification}. The horizontal axes 
    are velocities in $\kmps$, and the vertical axes are normalised
    flux.}
  
  \label{fig:atlas_reipurth}
  
\end{figure}

\citet{reipurth:1996} proposed a two-dimensional classification
of H$\alpha$ emission profiles of T~Tauri stars and Herbig Ae/Be
stars. Their classification scheme contains four classes (I, II, III
and IV) differentiated by the ratio of the secondary-to-primary emission
components in the profiles. Each class is divided into two sub-classes
(B and R) which depends whether the secondary peak is on the
blue or red side of the primary peak. Readers are referred to Fig.~4 of their paper.
Fig.~\ref{fig:atlas_reipurth} shows  sample model profiles
which are classified according to the definition of \citet{reipurth:1996}.
The combination of the disc wind, magnetospheric accretion flow, and
the accretion disc can reproduce all the classes of the profiles
seen in observations. The corresponding parameters of the
disc-wind-magnetosphere hybrid model used to reproduce the profiles
in the figure are summarised in Table~\ref{tab:classification} along
with brief comments on possible physical conditions which lead to
the profiles in each class. Below we discuss the Reipurth scheme in the
context of our models parameters. Readers are referred to
Appendix~\ref{sec:appendix} for the complete model profiles used in 
this discussion.          
Although in general H$\alpha$ in the spectra of CTTS shows a
significant amount of variablity, which may in some cases be induced by
the rotational motion of an inclined magnetosphere
(e.g.~\citealt{johns:1995}; \citealt{bouvier:2003}), we restrict our
discussion in this section to the axi-symmetric case only.

\textbf{Type~I}: 
\citet{reipurth:1996} found that Type I profiles (symmetric around
line centres) constituted 26~per~cent of the 43 CTTS H$\alpha$
profiles in their sample, making it the second most common
morphological type. We find that Type~I profiles are dominated by
magnetospheric emission at rates in excess of $\approx 10^{-8}\,
\mathrm{M_{\sun}\, yr^{-1}}$. Type~I profiles appear at a wide range
of inclinations (Fig.~\ref{fig:acc_inclination_effect}).  An object
that displays a Type~I profile may have 
relatively weak disc wind, but in this case it must be viewed from the
pole (in order that no wind material is seen projected against the
stellar photosphere).

\textbf{Type~II-B}:  
These profiles have a secondary blue peak that is in excess of half
the strength of the primary peak, and comprised 16~per~cent of the
Reipurth sample.
We find that Type~II-B profiles only occur at medium to high
inclinations, and for systems with mass accretion rates of $\approx
10^{-7}\, \mathrm{M_{\sun}\, yr^{-1}}$. If the system is viewed at
high inclination, a Type~II-B profile implies a fast wind
acceleration, and conversely a moderate inclination angle suggests a
more modest wind acceleration.

\textbf{Type~II-R}:  
These profiles are characterised by a secondary red peak that is in
excess of half the strength of the primary peak. They are
approximately as common as the Type~II-B lines.
For the $\dot{M}_{{\rm acc}} = 10^{-7}\,
\mathrm{M_{\sun}\,yr^{-1}}$ a switch from Type~II-B profiles to
Type~II-R occurs when the inclination angle changes from $55^\circ$ to
$80^\circ$. This profile type is also seen at lower mass-accretion
rates, but typically only in the models with the slowest accelerating
winds. It is perhaps unsurprising that the Type~II-B and Type~II-R
profiles are equally common, since (in general) the split between the
profile types results from a geometrical (viewing angle) effect rather
than a marked difference in physical parameters (mass accretion rate,
wind acceleration etc).
				   
\textbf{Type~III-B}:
Profiles with a blue secondary peak that is less than half the
strength of the primary peak are classified as Type~III-B. This type
of profile is the most common in the Reipurth sample, comprising 14
out of 43 (33 per cent) CTTS.
We find Type~III-B profiles only in our $\dot{M}_{{\rm acc}} = 10^{-7}\,
\mathrm{M_{\sun}\,yr^{-1}}$ models. A fast wind acceleration and
a moderate inclination are necessary, or a slower accelerating wind
viewed pole-on.  We do not see this profile type in our high
inclination models.
Although Type~III-B profiles do not occur at
$\dot{M}_{\mathrm{acc}} =
1.0\times10^{-8}\,\mathrm{M_{\sun}\,yr^{-1}}$, we have examined
the intermediate case of
$\dot{M}_{\mathrm{acc}} =3\times10^{-8}\,\mathrm{M_{\sun}\,yr^{-1}}$, which may
represent a typical T Tauri mass accretion rate
(\citealt{gullbring:1998}; \citealt{calvet:2004}). We find that
the models with fast accelerating, high temperature winds do
display Type~III-B profiles. It is clear that our model, as it
stands, requires above-average (but perhaps not atypical)
mass-accretion rates in order to produce Type~III-B profiles. It
is less clear whether the 14 Type~III-B objects found by
\citet{reipurth:1996} constitute the highest mass-accretion rate
objects in their sample: the average EW for T~Tauri stars with
Type~III-B profiles is 64~\AA, which is in the top quartile of
their complete T~Tauri star sample, perhaps indicating that
Type~III-B profiles are associated with high EWs, and therefore
strong mass-accretion.

\textbf{Type~III-R}: 
This profile type has a red secondary peak that is less than half the
strength of the primary peak, and only one object with this type of
profile was observed in the Reipurth sample (SZ Cha).
We are able to reproduce this profile type, but only using a very
narrow range of parameter space. In fact there are no Type~III-R
profiles in our model grid, but we obtain one by `tweaking' the magnetospheric
temperature, and using an inclination ($85^\circ$) sufficient that the
disc starts to obscure the central H$\alpha$ emission. For
completeness, the final model parameters we used to achieve the
profiles are: very high inclination,
$T_{\mathrm{max}}\sim8300\,\mathrm{K}$, and
$\dot{M}_{\mathrm{acc}}\sim10^{-8}\,\mathrm{M_{\sun}yr^{-1}}$.
It is interesting that the most infrequently observed profile type was
in fact the most difficult for us to reproduce. It appears that this
profile morphology requires some obscuration effects by the dust disc,
and therefore a high inclination; this naturally explains the rarity
of the profile in the observations.

\textbf{Type~IV-B}: 
This profile classification is the classical P-Cygni, in which the
blue-shifted absorption component has a sufficient velocity to be
present beyond the emission line wing. It occurs infrequently in
observations (7 per cent of the Reipurth sample).
No Type~IV-B profiles were produced in our model grid, since the blue
wing of the magnetosphere has generally such a large extent that it
always exceeds the projected wind terminal velocity, and thus the
bluemost extent of the absorption component. Only by running a model
with a larger terminal velocity or with a smaller magnetospheric
temperature, were we able to achieve a classical
P-Cygni shape. We suspect that the disc-wind geometry is not the best
kinematic model for producing this morphology. A more straightforward
explanation lies with a bipolar flow, viewed along its axis, or even a
simple spherical wind.

\textbf{Type~IV-R}: 
This is the inverse P-Cygni profile type, and only occurs in about
5~per~cent of H$\alpha$ profiles.
Type~IV-R profiles occur very infrequently in our hybrid model grid,
with only the lowest mass-accretion rate models viewed at high
inclination showing the characteristic shape (although using the EW
criterion these models would not be classified as CTTS). The inverse
P-Cygni profile requires that the observer view the hot accretion
footprints through the fast moving magnetospheric accretion flow, and
this means that the inclination angle is restricted to small range.
We can reproduce Type~IV-R profiles more readily if we adopt a cooler
magnetospheric temperature than indicated by the
accretion-rate/temperature relationship presented by
\citet{muzerolle:2001}. 
The restriction in the inclination angle mentioned above would be
relaxed if the axis of the magnetosphere was tilted with respect to the
rotational axis since the line-of-sight from an observer to the hot
rings will pass through  different parts of the 
magnetosphere as the rotational phase changes.

\begin{table*}

\caption{The summary of model parameters for the profiles in
Fig.~\ref{fig:atlas_reipurth} and brief comments. The temperatures
are in $10^{3}\,\Kelvin$. $\dot{M}_{\mathrm{acc}}$ and
$\dot{M}_{\mathrm{wind}}$ are in $\mathrm{M_{\sun}\,yr^{-1}}$.}

\label{tab:classification}

\begin{center}

\begin{tabular}{llllllll}
\hline 
Class&
$i$&
$\dot{M}_{\mathrm{acc}}$&
$T_{\mathrm{max}}$&
$\dot{M}_{\mathrm{wind}}$&
$T_{\mathrm{wind}}$&
$\beta$&
Comment\tabularnewline
\hline 
I&
$55^{\circ}$&
$10^{-7}$&
$7.5$&
$-$&
$-$&
$-$&
Accretion dominated. Wide range of inclination.\tabularnewline
II-B&
$55^{\circ}$&
$10^{-7}$&
$7.5$&
$10^{-8}$&
$8.0$&
$1.0$&
Wide range of wind acceleration rates.
Mid--high inclination.\tabularnewline
II-R&
$80^{\circ}$&
$10^{-7}$&
$7.5$&
$10^{-8}$&
$7.0$&
$2.0$&
Slow wind acceleration rate. High inclination.\tabularnewline
III-B&
$55^{\circ}$&
$10^{-7}$&
$7.5$&
$10^{-8}$&
$9.0$&
$0.5$&
Fast wind acceleration rate. Mid inclination.\tabularnewline
III-R&
$85^{\circ}$&
$10^{-8}$&
$8.2$&
$10^{-9}$&
$6.0$&
$1.0$&
Mid wind acceleration rate. Very high
inclination.\tabularnewline
IV-B&
$30^{\circ}$&
$10^{-7}$&
$5.5$&
$10^{-8}$&
$6.0$&
$0.2$&
Fast wind acceleration rate. Mid--high inclination.\tabularnewline
IV-R&
$55^{\circ}$&
$10^{-9}$&
$9.5$&
$-$&
$-$&
$-$&
Accretion dominated. Low mass-accretion rate. Mid inclination.\tabularnewline
\hline
\end{tabular}

\end{center}

\end{table*}

%
%
%

\subsection{The inclination dependence of H$\alpha$ profiles}

\label{sub:ha-mass-accretion-diagnostic}

\begin{figure*}
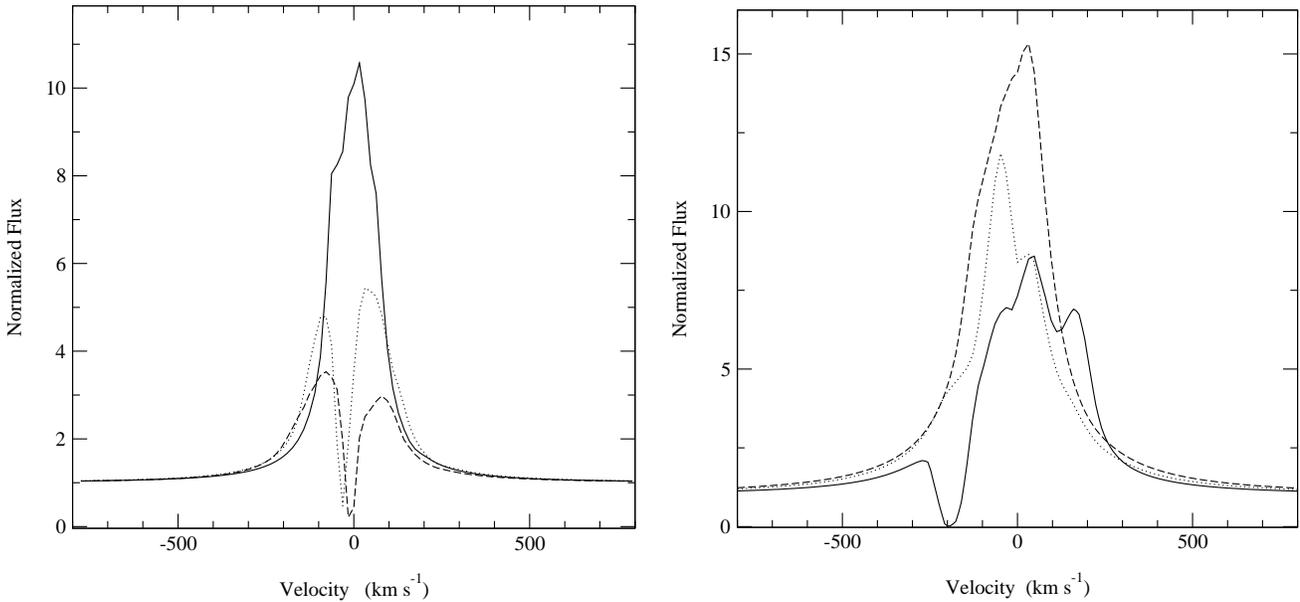


  \begin{center}

    \begin{tabular}{cc}
      \includegraphics[%
	clip,
	scale=0.45]{fig13a.eps}&
      \includegraphics[%
	clip,
	scale=0.61]{fig13b.eps}\tabularnewline
    \end{tabular}
    
  \end{center}

  \caption{The dependence of the H$\alpha$ model profiles on the
    inclination angle. The profiles computed at $i=10^{\circ}$ (solid
    line), $55^{\circ}$ (dotted line) and $80^{\circ}$ (dashed line)
    using (1)~the disc-wind-magnetosphere hybrid model (left;
    c.f. Section~\ref{sub:result-discwind-knigge}) and (2)~the disc,
    stellar wind, magnetosphere hybrid model (right panel;
    c.f. Fig.~\ref{fig:config_stellarwind}).  The parameters used for
    the magnetosphere are same for the both models,
    i.e. $T_{\mathrm{max}}=7500~\mathrm{K}$ and
    $\dot{M}_{\mathrm{acc}}=10^{-7}~\mathrm{M_{\sun}~yr^{-1}}$.  The
    isothermal wind temperature and the wind acceleration parameter
    used in (1) are $T_{\mathrm{wind}}=7000~\mathrm{K}$ and
    $\beta=1.0$, and those in (2) are
    $T_{\mathrm{wind}}=9000~\mathrm{K}$ and $\beta=1.0$ respectively.
    The wind mass-loss rate used in both models is
    $\dot{M}_{\mathrm{wind}}=10^{-8}~\mathrm{M_{\sun}~yr^{-1}}$ . 
    All the other parameters are same as in
    Sections~\ref{sub:result-discwind-hybrid} and
    \ref{sub:result-discwind-knigge}.  }

  \label{fig:effect_inclination_compare}
\end{figure*}

Recently, \citet{appenzeller:2005} have shown that the EW of the
observed H$\alpha$ from CTTS is inclination dependent, with the EW
increasing as the inclination angle increases. Next, we will examine
whether the hybrid model is consistent with the inclination dependency
found in the observation.
The left-hand panel in Fig.~\ref{fig:effect_inclination_compare}
shows the profiles computed with the disc-wind, magnetosphere hybrid
model (Section~\ref{sub:result-discwind-knigge}) at inclination angles
$i=10^{\circ}$, $55^{\circ}$ and $80^{\circ}$. The magnetospheric
accretion component uses the parameters used as the reference model
(Table~\ref{tab:std_parameters}).  The figure shows that the
absorption feature becomes stronger as the inclination increases,
mainly because of the geometrical configuration.  The optical depth to
the observer becomes larger as the inclination angle increases, since
the density of the disc wind increases towards the equatorial
plane -- as a result, the EW becomes smaller as the inclination angle
increases: 41, 27, and 19\,\AA\, for $i=10^\circ$,
$55^\circ$, and $80^\circ$ respectively.  This is a robust result, and
is insensitive to the adopted wind parameters. Recall that this is the
same EW/inclination dependency as the magnetosphere only models
(Fig.~\ref{fig:acc_inclination_effect}).

Clearly the inclination dependency of the line strength is in
contradiction with the \citet{appenzeller:2005} result, and thus
merits further investigation.  In the previous sections, we have
considered H$\alpha$ line formation in a disc wind, which mimics the
density distribution of the magneto-centrifugal launched jet models
(e.g.~\citealt{krasnopolsky:2003}). The density in this model is more
concentrated toward the equatorial plane in the region where the line
emissivity is highest, although further out the material becomes
collimated into a bipolar jet, this would not contribute significantly
to the line emission in our models.

\begin{figure}
  
  \begin{center}
    
    \includegraphics[%
      scale=0.3]{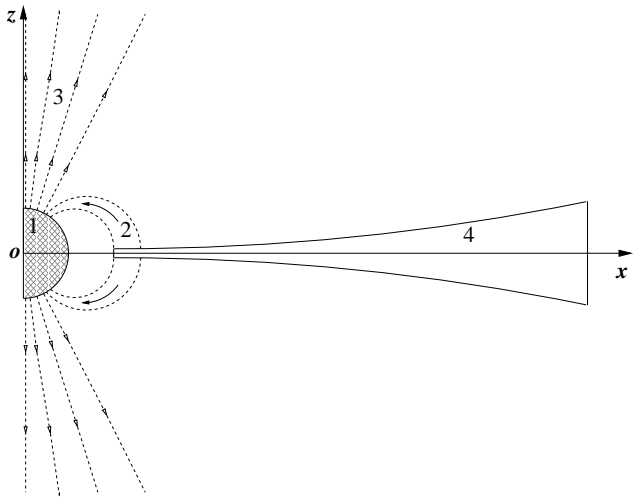}
    
  \end{center}
  
  \caption{The alternative outflow model. This model replaces the
    disc-wind with a stellar wind. The
    wind is launched from the stellar surface, and is limited to 
    cones with opening angles of $\sim30^\circ$.  The system consists of four
    components: (1)~the continuum source located at the origin $\left( o
    \right)$ of the cartesian coordinates $\left(x,y,z\right)$ -- the
    $y$-axis is into the paper, (2)~the magnetospheric accretion flow,
    (3)~the stellar wind outflow, and (4)~the accretion disc. The density
    distribution is symmetric around the $z$-axis. }
  
  \label{fig:config_stellarwind}
  
\end{figure}

As an alternative, we consider a wind model in which the outflow
density increases towards the pole, rather than the equator. In this
model, the disc-wind is replaced by a stellar wind, propagating only
in the radial direction, accelerating under the classical
beta-velocity law \citep[c.f.][]{castor:1979} i.e.
$v_{r}\left(\ r  \right) = v_{\infty} \left( 1 - R_{*}/r
  \right)^{\beta}$  where  $v_{\infty}$ and  $R_{*}$ are the 
  wind terminal velocity and the stellar radius respectively. $\beta$
  is the wind acceleration parameter similar to the one seen in the
  disc-wind model (c.f.~equation~\ref{eq:discwind-poloidal-velocity}).
The wind emerges from the stellar surface, but restricted to be within
$\sim30^{\circ}$ from the symmetry axis hence forming the cone-shaped 
regions both above and below the poles. The density is computed from
assuming a constant mass-loss rate per unit area over the stellar
surface, requiring mass-flux conservation, and using the beta velocity law
above.  The schematic of the model configuration is shown in
Fig.~\ref{fig:config_stellarwind}. Again, we assume the wind
temperature ($T_{\mathrm{wind}}$) is isothermal for simplicity.  
This model configuration is motivated by the recent study by
\citet{matt:2005} who demonstrated that a stellar
wind along  open magnetic field lines originating from the star may cause
significant spin-down torque.				   

The right-hand panel in Fig.~\ref{fig:effect_inclination_compare}
shows H$\alpha$ profiles computed using the alternative model at the
same inclination angles used for the disc-wind-magnetosphere hybrid
model. The mass-loss rate of the stellar wind in this model is
$10^{-8}\, \mathrm{M_{\sun}\,yr^{-1}}$, which
is the same as the mass-loss rate used in the disc-wind magnetosphere
hybrid model in the same figure.   In both models, the mass-accretion
rate of $10^{-7}\, \mathrm{M_{\sun}\,yr^{-1}}$ is used.
The wind acceleration parameter ($\beta$) and the wind
temperature ($T_{\mathrm{wind}}$) used here are $1.0$ and
$9000\,\Kelvin$ respectively. All the other parameters are the same as in
the disc-wind magnetosphere model.  With this set of $\beta$ and
$T_{\mathrm{wind}}$, the wind emission is significantly larger than
the emission from the magnetosphere.  Because of the geometry of the
wind, the P-Cygni absorption feature weakens as the inclination
increases: the optical depth of the wind is much higher in the polar
direction. For the same reason, the EW of the line increases as the
inclination increases, as clearly seen in the figure. Specifically the
values are 54, 69, and 98\,\AA\, for $i=10^\circ$,
$55^\circ$, and $80^\circ$ respectively.  
This inclination dependency of EW is seen in the models with wide ranges of 
$\beta$ and $T_{\mathrm{wind}}$, provided that the wind emission
of the model is dominating the emission from magnetosphere.
A similar inclination dependency of EW is found for lower wind
mass-loss rate models
(e.g.~$\dot{M}_{\mathrm{wind}}=10^{-9}\,\mathrm{M_{\sun}\,yr^{-1}}$)
while keeping $\mu=0.1$.

The \citet{appenzeller:2005} study used just 12 objects, and a significantly
greater sample should be investigated before firm  conclusions are
drawn. However our models indicate a significant difference in the 
inclination dependence of the line morphology which may be simply
tested against observation.  For example, one
possible way to distinguish the two different wind models is to 
examine the profile shape of a CTTS which is known to be viewed close
to pole-on (e.g. a CTTS with a very low $v \sin i$). 
If the profile contains a prominent blue-shifted absorption
component (e.g. Type IV-B), the system is more likely to be associated
with the bipolar stellar wind type of outflow, and not with the disc-wind.
In reality, however, the two types of winds may be present in the same 
object, in a similar fashion to that seen in the models of 
\citet*{drew:1998} and \citet{matt:2005}. In these models, the
fast stellar wind is present in the polar directions, while at the same
time a slower, denser disc-wind is present near the equatorial plane.

\section{Conclusions}

\label{sec:Conclusions}

We have presented disc-wind-magnetosphere hybrid radiative
transfer models for classical T~Tauri stars, and detailed studies of
the H$\alpha$ formation from their complex circumstellar
environment.  We found that the hybrid model can reproduce the wide
variety of profile seen in observations
(Figures~\ref{fig:atlas_acc_models}, \ref{fig:atlas_discwind_only},
\ref{fig:atlas_hybrid} and \ref{fig:atlas_reipurth}).

Using the model results, we examined the H$\alpha$ spectroscopic
classification proposed by \citet{reipurth:1996}, and discussed the
basic physical conditions that reproduce the profiles in each
classified type
(Section~\ref{sub:Classification-scheme-proposed}). Using the different
combinations of the inclination 
($i$), the mass-loss to mass-accretion rate ratio ($\mu$) and the wind
acceleration rate ($\beta$), our radiative transfer model was able to
produce all 7 classes of profiles defined in \citet{reipurth:1996}.

One potential free parameter of our models is the ratio of mass-loss
rate to mass-accretion rate ($\mu$). Most MHD simulations predict a
value of $\mu = 0.1$ (c.f.~\citealt{koenigl:2000}), and we have
adopted this figure for most of our models. The success of our profile
in matching observations is therefore encouraging, and provides
support for this canonical value of $\mu$. We have computed sample
models with other values of $\mu$,
(Fig.~\ref{fig:discwind_mu_effect}) and find that models with $0.05
<\mu<0.2$ are still consistent with observations. Wind-only or
magnetosphere-only models may be representative of individual line
profiles, but neither can explain the complete Reipurth classification
scheme -- even magnetosphere dominated profiles (Type~I) can be readily
explained by the hybrid model with $\mu=0.1$.

Unfortunately, the dependency of the line equivalent width on
the inclination angle predicted by  
the disc-wind-magnetosphere hybrid model does not agree with the trend
seen the observations of \citet{appenzeller:2005}. We may appeal
to the small sample size of their study (12 objects) to question
the validity of the EW effect, but we have also
considered an alternative  model in which the disc-wind is
replaced by a biconical stellar wind, and have found that this
alternative wind model can reproduce the 
line equivalent width dependency on the inclination 
(Fig.~\ref{fig:effect_inclination_compare}).  However, MHD simulations
predict that bipolar or jet-like flows in CTTS are always accompanied by
a disc-wind originating near the star: the actual situation therefore is likely
to be a combination of the two flows.

Although we have concentrated on CTTS here, the hybrid model is
equally applicable to pre-main-sequence stars across the mass
spectrum. Herbig Ae/Be star profiles are equally diverse, and whilst
many of the profiles are indicative of a wind
(\citealt{finkenzeller:1984}) some lines may be explained by
magnetospheric accretion alone (e.g. \citealt{muzerolle:2005}). One
possible advantage in applying the hybrid model to the Herbig Ae/Be
systems is the growing observational data probing the line formation
region and inner disc using techniques such as spectro-astrometry
\citep[e.g.][]{baines:2006}, spectropolarimetry
\citep[e.g.][]{vink:2005} and interferometry
\citep[e.g.][]{monnier:2005}. These high-resolution techniques may aid
model fitting of individual objects by constraining some of the model
parameters, in particular the inclination.

Velocity broadened, asymmetric H$\alpha$ profiles in brown dwarfs
(BDs) are used as an indicator of ongoing magnetospheric accretion,
and the same models that are used for CTTS are applied to these
systems in order to determine mass-accretion rates
(e.g.~\citealt*{lawson:2004}; \citealt{muzerolle:2005}). The line
profiles however display the same wide variety of morphologies as the
CTTS and Herbig Ae/Be stars (although we note as yet no BD profile has
been observed with red-shifted absorption).  It is natural to assume
that the same physics applies in these low mass objects, and that a
hybrid model may be more successful in fitting these profiles, and
will give greater insight into the circumstellar kinematics.

There are several avenues of future work to consider. Here we have
adopted a simple analytical form for the accretion and outflow
velocity and density structure.  A more self-consistent approach would
be to use the results of MHD simulations as an input to the
radiative-transfer modelling. The thermodynamics of the circumstellar
gas needs to be examined (c.f.~\citealt{hartmann:1994};
\citealt{martin:1996}) and also computed self-consistently with the
radiation field used in the radiative-transfer, a process that would
decrease the free parameters of the model.  We intend to calculate
other observables from the model that may be compared with observation
and should provide further constraints. The line polarisation
signature, which arises from scattering of line and continuum photons
in the circumstellar dust, encodes both dynamical and geometric
information, such as the size of the inner disk radius
(e.g.~\citealt{vink:2005a}: \citealt{vink:2005}). Spectro-astrometry
may provide information on the system's inclination, and the geometry
of the outflow on large scales (e.g.~\citealt{takami:2003}:
\citealt{baines:2006}). Interferometry also provides information on
inner disc radii and inclinations, although currently only for the
Herbig Ae/Be stars. 

We have concentrated on the formation of the H$\alpha$ line, because it
is the best studied CTTS emission line. However it is clear that
simultaneous fitting of several emission line profiles should provide
much stronger constraints. In particular it appears that the near-IR
hydrogen lines (Pa$\beta$ and Br$\gamma$) are less affected by
outflows than H$\alpha$. One could therefore use these lines to
constrain the mass-accretion rate, and use H$\alpha$ to obtain the
mass-loss rate. A potential pitfall is the significant temporal
variability that all CTTS show in their emission lines; the
observational data on a particular object must be co-temporaneous.

\section*{Acknowledgements}

We thank the referee, Silvia Alencar, who provided us with valuable comments and
suggestions which improved the clarity of the manuscript. This work
was supported by PPARC standard grant PPA/G/S/2001/00081 and PPARC
rolling grant PP/C501609/1.

%
%


%
%

%
%
\appendix

\section{H$\alpha$ atlas of the disc-wind-magnetosphere hybrid model}
\label{sec:appendix}

\begin{figure}
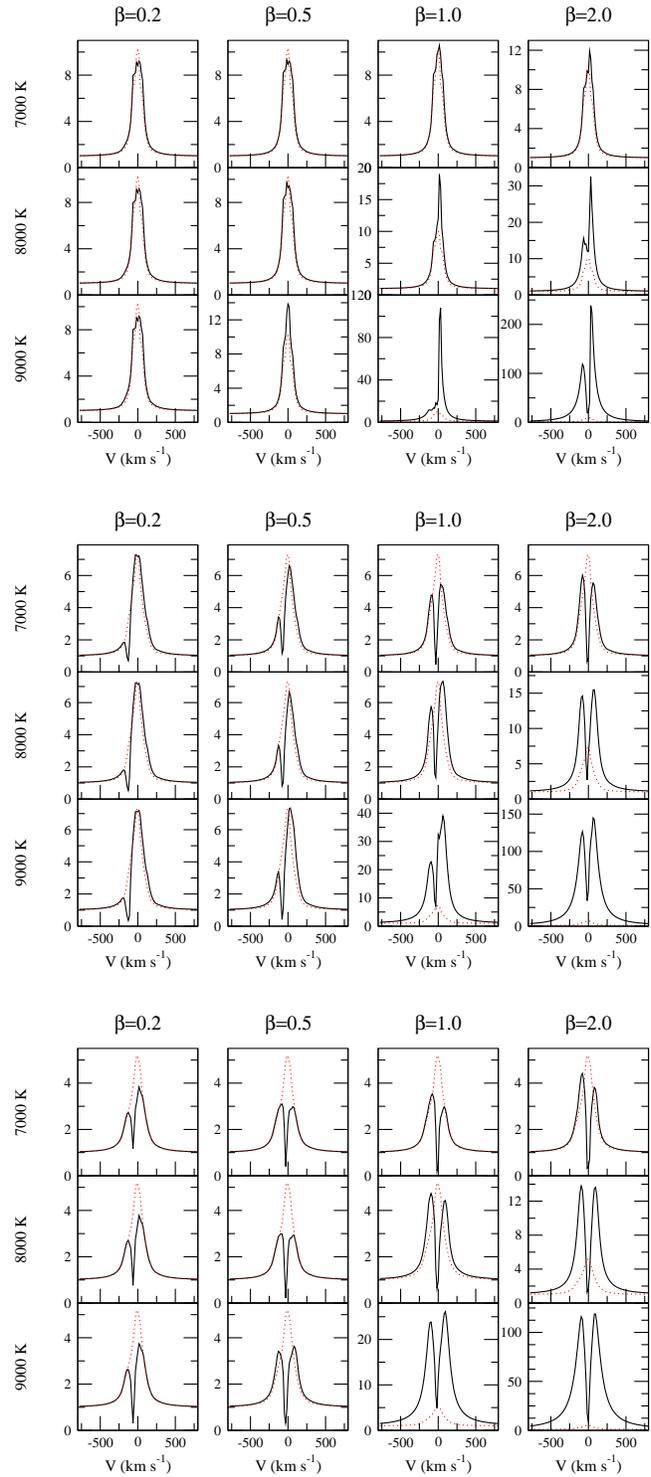


  \begin{center}
      \includegraphics[%
	clip,
	scale=0.6]{fig_a1_01.eps}\tabularnewline
      \vspace{0.5 cm}

      \includegraphics[%
	clip,
      scale=0.6]{fig_a1_02.eps}\tabularnewline

      \vspace{0.5 cm}

      \includegraphics[%
	clip, scale=0.6]{fig_a1_03.eps}\tabularnewline
  \end{center}

  \caption{An H$\alpha$ atlas of the disc-wind-magnetosphere hybrid
  model with $\dot{M}_{\mathrm{acc}}
  =10^{-7}\,\mathrm{M_{\sun}\,yr^{-1}}$ and
  $\dot{M}_{\mathrm{wind}}=10^{-8}\,\mathrm{M_{\sun}\,yr^{-1}}$.
  Profiles are presented for inclinations of $10^\circ$ (top figure),
  $55^\circ$ (middle figure), and $80^\circ$ (bottom figure).  For
  each inclination, 12 profiles are given, for three different values
  of the wind temperature, and four different values of the wind
  acceleration parameter $\beta$. The contribution of the
  magnetosphere can be judged from the profiles computed with the
  magnetosphere only (dotted lines).}

  \label{fig:atlas01}

\end{figure}

\begin{figure}

  \begin{center}
      \includegraphics[%
	clip,
	scale=0.6]{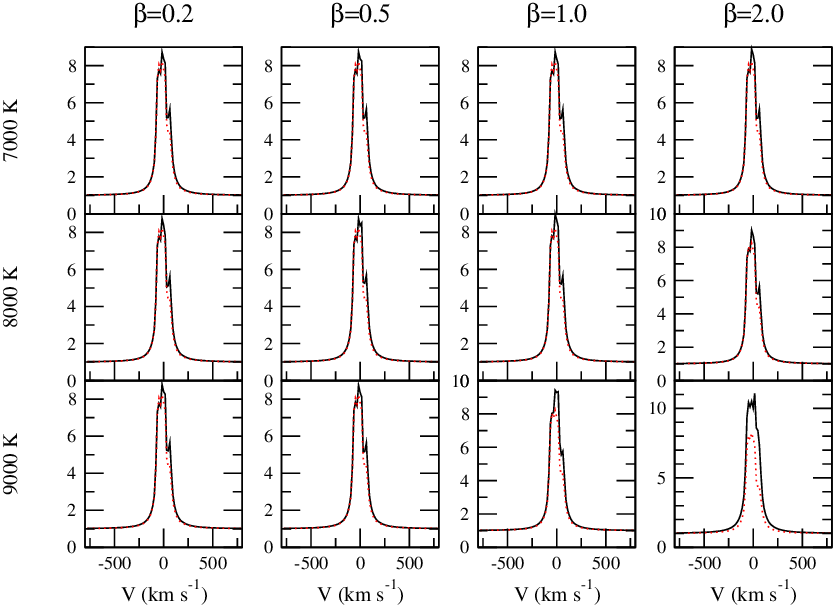}\tabularnewline
      \vspace{0.5 cm}

      \includegraphics[%
	clip,
      scale=0.6]{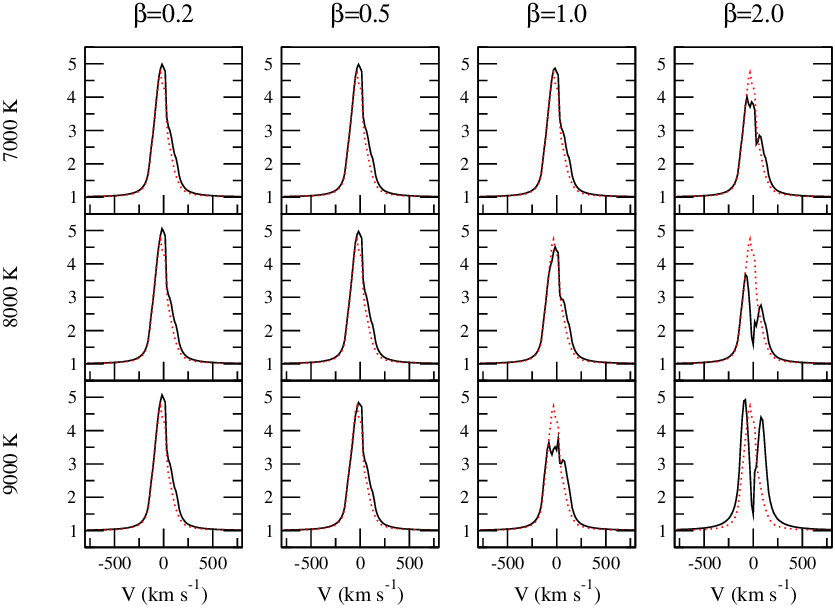}\tabularnewline

      \vspace{0.5 cm}

      \includegraphics[%
	clip, scale=0.6]{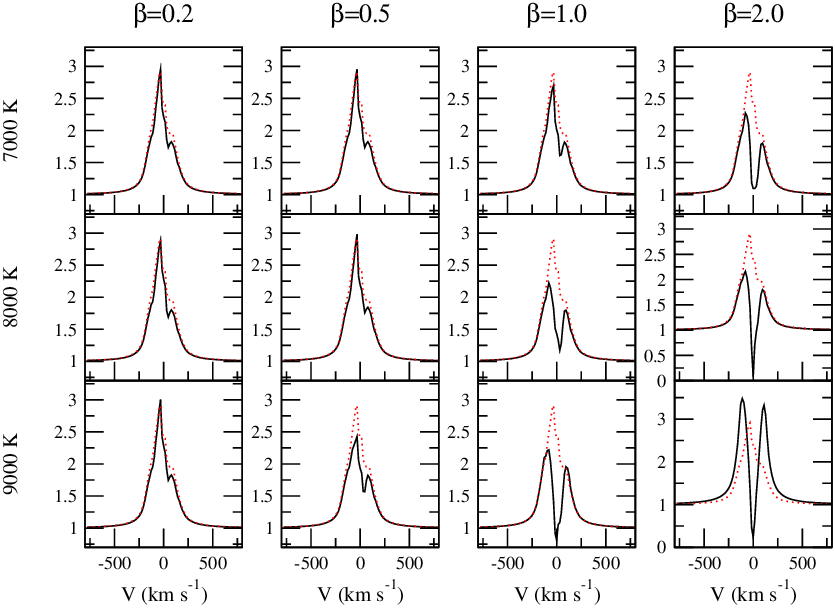}\tabularnewline
  \end{center}

  \caption{As for Fig.~\ref{fig:atlas01} except
  $\dot{M}_{\mathrm{acc}} =10^{-8}\,\mathrm{M_{\sun}\,yr^{-1}}$ and
  $\dot{M}_{\mathrm{wind}}=10^{-9}\,\mathrm{M_{\sun}\,yr^{-1}}$. } 

  \label{fig:atlas02}

\end{figure}

\begin{figure}

  \begin{center}
      \includegraphics[%
	clip,
	scale=0.6]{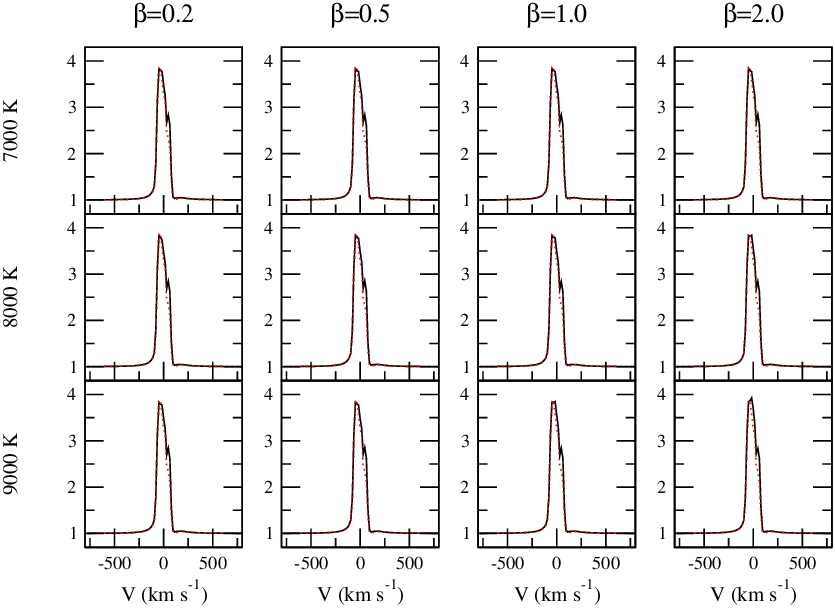}\tabularnewline
      \vspace{0.5 cm}

      \includegraphics[%
	clip,
      scale=0.6]{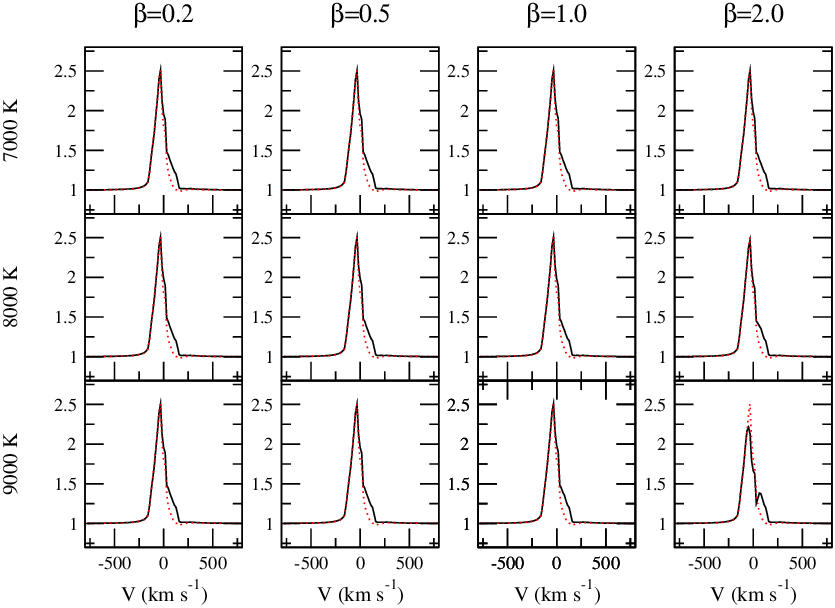}\tabularnewline

      \vspace{0.5 cm}

      \includegraphics[%
	clip, scale=0.6]{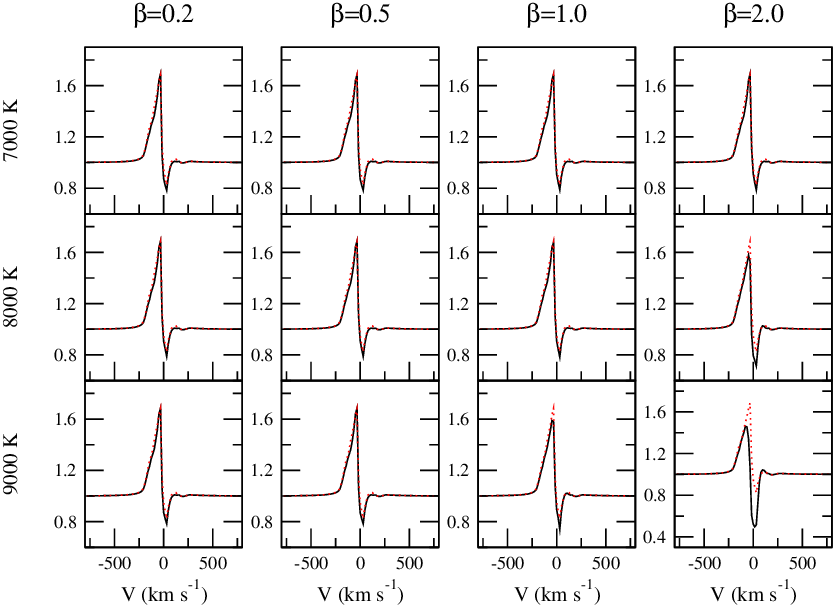}\tabularnewline
  \end{center}

  \caption{As for Fig.~\ref{fig:atlas01} except
    $\dot{M}_{\mathrm{acc}} =10^{-9}\,\mathrm{M_{\sun}\,yr^{-1}}$ and
  $\dot{M}_{\mathrm{wind}}=10^{-10}\,\mathrm{M_{\sun}\,yr^{-1}}$. } 

  \label{fig:atlas03}

\end{figure}

Here we present our H$\alpha$ model line profile grid 
computed using the disc-wind-magnetosphere hybrid model
 in Section~\ref{sub:result-discwind-hybrid}.  All
the models use the reference parameters for the 
magnetospheric geometry and the photosphere as given in
Table~\ref{tab:std_parameters} in the main text. The ratio ($\mu$) of the
mass-loss rate to mass-accretion rate is fixed to 0.1 for  all the 
models. Three sets of model profiles, each with a different
mass-accretion rate, are presented here: 
(1) $\dot{M}_{\mathrm{acc}}=10^{-7}\,\mathrm{M_{\sun}\,yr^{-1}}$ (Fig.~\ref{fig:atlas01}), 
(2) $\dot{M}_{\mathrm{acc}}=10^{-8}\,\mathrm{M_{\sun}\,yr^{-1}}$ (Fig.~\ref{fig:atlas02}), and
(3) $\dot{M}_{\mathrm{acc}}=10^{-9}\,\mathrm{M_{\sun}\,yr^{-1}}$
(Fig.~\ref{fig:atlas03}).   The temperatures of the magnetospheres
used for (1), (2) and (3) are $7500\,\Kelvin$, $8500\,\Kelvin$ and
$9500\,\Kelvin$ respectively. The profiles are presented for the
inclination angles of $10^\circ$, $55^\circ$, and $80^\circ$. 
The disc-wind geometry used to compute these profiles are the same as in
Section~\ref{sub:result-discwind-hybrid}. These profiles are available
electrically at \url{http://www.astro.ex.ac.uk/people/rk/profiles}.

\label{lastpage}
\end{document}